\documentclass[
aps,
prb,
reprint,
superscriptaddress,
nofootinbib,
longbibliography,
floatfix
]{revtex4-2}

\usepackage{amsmath,amssymb,bm}
\usepackage{graphicx}
\usepackage{booktabs}
\usepackage[hidelinks]{hyperref}

\newcommand{\dR}{\delta R}
\newcommand{\Rbar}{\bar{R}}
\newcommand{\dA}{\Delta A}
\newcommand{\Qs}{{e_s}}
\newcommand{\Nn}{{e_u}}

\begin{document}

\title{Antisymmetric breathing in altermagnetic skyrmions}

\author{Yunxi Jiang}
\affiliation{Department of Physics, Xi'an Jiaotong-Liverpool University, 215123, Suzhou, China}
\affiliation{Universit\'e de Strasbourg, CNRS, Institut de Physique et Chimie des Mat\'eriaux de Strasbourg, UMR 7504, F-67000 Strasbourg, France}
\author{Chen Xuan}
\affiliation{Department of Applied Mathematics, Xi'an Jiaotong-Liverpool University, 215123, Suzhou, China}
\author{Xi Chen}
\affiliation{Department of Physics, Xi'an Jiaotong-Liverpool University, 215123, Suzhou, China}
\author{Zhikai Wang}
\affiliation{Department of Physics, Xi'an Jiaotong-Liverpool University, 215123, Suzhou, China}
\author{Qinfeng Zhu}
\affiliation{Department of Computer Science, University of Liverpool, Liverpool L69 3BX, UK}
\author{Hao Yu}
\email{Hao.Yu@xjtlu.edu.cn}
\affiliation{Department of Physics, Xi'an Jiaotong-Liverpool University, 215123, Suzhou, China}

\date{\today}

\begin{abstract}
A skyrmion in an altermagnet with \(d\) wave symmetry consists of two elliptical sublattice textures with perpendicular long axes.
For each sublattice component $\eta=A,B$, we define an effective skyrmion radius $R_\eta=\sqrt{a_\eta b_\eta}$, where $a_\eta$ and $b_\eta$ are the distances from the skyrmion center to its boundary along $y$ and $x$, respectively.
For the anisotropic exchange parameters studied, the equilibrium radius at zero field is smaller than in a reference with parallel sublattice textures and otherwise identical parameters.
A magnetic field perpendicular to the film expands one sublattice skyrmion and contracts the other, generating a radius difference $\dR=R_A-R_B$ and a net magnetic moment.
The exchange modulation used to represent uniaxial strain also produces a nonzero radius difference at zero field.
Because the strong and weak exchange directions are interchanged between the two sublattices, a common directional change of the exchange couplings increases one radius and decreases the other.
Unlike the magnetic coupling, this mechanism vanishes when the two sublattice exchange tensors become identical.
The radius difference also supports an antisymmetric breathing mode.
After a short field pulse, it oscillates in quadrature with the uniform helicity, the common rotation of the wall magnetization within the film plane, at \(49.5\,\mathrm{GHz}\) for the reference parameters.
\end{abstract}

\maketitle

\section{Introduction}
Altermagnets have two spin sublattices which are related by a crystal rotation rather than by translation or inversion \cite{PhysRevX.12.040501,PhysRevX.12.031042,Song2025NRM,v867-h742,1vqq-9kzm}.
Their momentum-dependent spin
splitting is affected by the magnetic crystal symmetry, and $d$-wave altermagnetism has been observed
in a metallic room-temperature material \cite{Jiang2025NP}.
In the $d$-wave model used here, the two sublattices carry interchanged in-plane exchange
stiffnesses, and the skyrmion consists of two elliptical sublattice textures whose long
axes are perpendicular \cite{PhysRevLett.133.196701,xtcd-t47t,PhysRevLett.134.176401,9vpq-hp7b}.
Skyrmions in antiferromagnets, ferrimagnets, ferromagnets, and magnetic multilayers could be manipulated by optical excitation, spin orbit torque, electric fields, and light induced strain \cite{PhysRevB.107.214429,Liu_2023,567,article345,articleer,https://doi.org/10.1002/adma.202270090}. 
For the altermagnet skyrmions,
previous works have examined their its magnetic multipole, anisotropic Hall response,
spin-transfer torques, current-driven dynamics, and interaction with pinning
\cite{PhysRevLett.133.196701,xtcd-t47t,PhysRevLett.134.176401,y9q4-13fw,vplq-sd5k,liu2026emergentskyrmionhalleffect}.

At zero field, the two sublattice skyrmions carry opposite topological charges, and their net magnetic moment vanishes.
The two perpendicular elliptical contours intersect along four directions.
Their local mismatch elsewhere produces an alternating four-lobed pattern.
This residual magnetization generates a magnetostatic field that acts on both sublattice textures.

For each sublattice component $\eta=A,B$, we define an effective skyrmion radius
$R_\eta=\sqrt{a_\eta b_\eta}$, where $a_\eta$ and $b_\eta$ are the distances from the skyrmion
center to its boundary along $y$ and $x$, respectively.
Our simulations show that an out-of-plane magnetic field enlarges one sublattice skyrmion and shrinks the other. 
This opposite change of the two radii motivates the antisymmetric breathing coordinate studied here.

Breathing modes of single and antiferromagnetic
skyrmions, as well as in-phase and antiphase modes of exchange-coupled skyrmions, are already
known \cite{PhysRevB.99.054430,PhysRevB.99.184429,PhysRevB.102.104403}.
In a \(d\)-wave altermagnet, the exchange-anisotropy axes are interchanged between the two sublattices, making the two elliptical skyrmions perpendicular. We study how this altermagnetic geometry changes the equilibrium size of the skyrmion.
To isolate the role of this orientation, we compare it with
a reference in which the two ellipses have parallel long axes.
The coordinate studied throughout the work is \(\dR=R_A-R_B\).
In this work, the skyrmion cores on sublattices \(A\) and \(B\) point along \(+z\) and \(-z\), respectively, so a \(+z\) field expands the \(A\) skyrmion and contracts the \(B\) skyrmion.
At zero field, where \(R_A=R_B\), the two perpendicular walls still do not coincide. 
The perpendicular geometry has a smaller radius than a parallel-ellipse reference built from the same exchange constants. 
We also generate a radius difference at zero field by applying a directional exchange modulation, which we use to represent uniaxial strain. We apply the same directional perturbation to the exchange couplings on both sublattices: all \(x\)-directed exchange coefficients are multiplied by \(1+\delta\), while all \(y\)-directed coefficients are multiplied by \(1-\delta\). Because the strong and weak exchange directions are interchanged between the two sublattices, the same perturbation makes one skyrmion more elongated and the other less elongated. As a result, a nonzero radius difference occurs.
Beyond these static changes, the radius difference coordinate also supports a dynamical mode. After a short out-of-plane field pulse, the radius difference oscillates together with the uniform helicity, defined as the common rotation of the in-plane magnetization in the two domain walls. The helicity reaches an extremum when the radius difference crosses zero.

\section{Model and simulated skyrmion}
\label{sec:model}
The lattice Hamiltonian for a $d$-wave altermagnet is
\cite{PhysRevLett.133.196701,xtcd-t47t},
\begin{equation}
\begin{aligned}
\mathcal{H}=\sum_{i,j}\Big[
&-J_1\,\mathbf{m}^A_{i,j}\!\cdot\!\mathbf{m}^A_{i+1,j}
 -J_2\,\mathbf{m}^B_{i,j}\!\cdot\!\mathbf{m}^B_{i+1,j}\\
&-J_2\,\mathbf{m}^A_{i,j}\!\cdot\!\mathbf{m}^A_{i,j+1}
 -J_1\,\mathbf{m}^B_{i,j}\!\cdot\!\mathbf{m}^B_{i,j+1}\\
&-J_3\,\mathbf{m}^A_{i,j}\!\cdot\!\mathbf{m}^B_{i,j}
 -g\mu_B\mathbf{B}\!\cdot\!(\mathbf{m}^A_{i,j}+\mathbf{m}^B_{i,j})\\
&-D_0\big(\mathbf{m}^A_{i,j}\!\times\!\mathbf{m}^A_{i+1,j}
        +\mathbf{m}^B_{i,j}\!\times\!\mathbf{m}^B_{i+1,j}\big)\!\cdot\!\hat{e}_y\\
&+D_0\big(\mathbf{m}^A_{i,j}\!\times\!\mathbf{m}^A_{i,j+1}
        +\mathbf{m}^B_{i,j}\!\times\!\mathbf{m}^B_{i,j+1}\big)\!\cdot\!\hat{e}_x\\
&+K_0\big[1-(\mathbf{m}^A_{i,j}\!\cdot\!\hat{e}_z)^2\big]
 +K_0\big[1-(\mathbf{m}^B_{i,j}\!\cdot\!\hat{e}_z)^2\big]\Big]\\
&+\mathcal H_{\rm ms}.
\end{aligned}
\label{eq:Ham}
\end{equation}
Here, $\mathbf m^\eta_{i,j}$ is the unit magnetization vector on sublattice $\eta=A,B$ at lattice site $(i,j)$. The parameters $J_1$ and $J_2$ are the nearest-neighbor exchange couplings within each sublattice. Their directions are interchanged between the two sublattices: sublattice $A$ has exchange couplings $J_1$ and $J_2$ along $x$ and $y$, respectively, whereas sublattice $B$ has $J_2$ and $J_1$.
The parameter $J_3<0$ describes the antiferromagnetic exchange coupling between the two sublattices.
The Zeeman term contains the Land\'e factor $g$, the Bohr magneton $\mu_B$, and the applied magnetic field $\mathbf B$.
$D_0$ is the strength of the interfacial Dzyaloshinskii--Moriya interaction (DMI) within each sublattice, and $K_0>0$ is the perpendicular easy-axis anisotropy along $z$.
The term $\mathcal H_{\rm ms}$ is the magnetostatic energy generated by the spatial distribution of the combined sublattice magnetization. The corresponding continuum energy density is
\begin{equation}
\begin{aligned}
w&=
A_1|\partial_x\mathbf m_A|^2
+A_2|\partial_y\mathbf m_A|^2
+A_2|\partial_x\mathbf m_B|^2
+A_1|\partial_y\mathbf m_B|^2\\
&+K(1-m_{Az}^2)
+K(1-m_{Bz}^2)\\
&+D\big(
m_{Az}\partial_x m_{Ax}
+m_{Az}\partial_y m_{Ay}
-m_{Ax}\partial_x m_{Az}\\
&-m_{Ay}\partial_y m_{Az}
\big)+D\big(
m_{Bz}\partial_x m_{Bx}
+m_{Bz}\partial_y m_{By}\\
&-m_{Bx}\partial_x m_{Bz}
-m_{By}\partial_y m_{Bz}
\big)\\
&-M_s\mathbf B\cdot\mathbf m_A
-M_s\mathbf B\cdot\mathbf m_B
+A_0\mathbf m_A\cdot\mathbf m_B\\
&+w_{\rm ms}.
\end{aligned}
\label{eq:w}
\end{equation}
The continuum coefficients are related to the lattice parameters by
$
A_{1,2}=\frac{J_{1,2}a^2}{2v_c},
A_0=-\frac{J_3}{v_c},
D=-\frac{D_0a}{v_c},
K=\frac{K_0}{v_c},
M_s=\frac{g\mu_B}{v_c}
$
where $v_c=a^2t_c$ is the unit-cell volume, $a$ the in-plane lattice constant, $t_c$ the
out-of-plane lattice period, and $M_s$ the saturation magnetization of each sublattice. 
The magnetostatic term is
\[
 w_{\rm ms}(\mathbf r)=-\frac12\mathbf M(\mathbf r)\cdot\mathbf H_{\rm ms}(\mathbf r),
 \qquad \mathbf M=M_s(\mathbf m_A+\mathbf m_B),
\]
where $\mathbf H_{\rm ms}$ is the demagnetizing field expressed in tesla, with $\mu_0$ included in its definition.
For a film of thickness $t$, the magnetostatic energy is
$E_{\rm ms}=t\int d^2r\,w_{\rm ms}$. The field at each point is generated by the total magnetization throughout the film and acts on both sublattices. The thickness-averaged magnetostatic kernel is given in Supplemental Material, Sec.~S1. The local anisotropy is
$K=K_0/v_c$.
The magnetization dynamics of each sublattice are described by the
Landau--Lifshitz--Gilbert (LLG) equation
\begin{equation}
\frac{\partial \mathbf m_\eta}{\partial t}
=
-\frac{\gamma}{1+\alpha^2}
\left[
\mathbf m_\eta\times \mathbf H_{\eta,\mathrm{eff}}
+
\alpha\,\mathbf m_\eta\times
\left(
\mathbf m_\eta\times\mathbf H_{\eta,\mathrm{eff}}
\right)
\right],
\label{eq:simulation_llg}
\end{equation}
where $\gamma=1.76\times10^{11}$ rad\,s$^{-1}$\,T$^{-1}$ and
$\alpha$ is the Gilbert damping parameter.
The effective fields, expressed in tesla, are $\mathbf H_{\eta,\mathrm{eff}}(\mathbf r)=-(M_st)^{-1}\delta E/\delta\mathbf m_\eta(\mathbf r)$, with $E=t\int d^2r\,w$.
We define
\begin{equation}
\bar{A}=\sqrt{A_1A_2},\qquad \dA=A_1-A_2.
\label{eq:Abar}
\end{equation}
Here the geometric mean $\bar A$ sets the overall exchange-stiffness scale, whereas $\Delta A$
measures the difference between the two in-plane directions. When $\Delta A=0$, the exchange is
isotropic and the exchange contrast of this two-sublattice model disappears.
We minimize the total energy $E$ in a $96\times96\,\mathrm{nm}^2$ periodic cell on a $192\times192$ grid with spacing $h=0.5\,\mathrm{nm}$. The minimization procedure and the extraction of radii and ellipticities are described in Supplemental Material, Sec.~S2.
Following the parameterization in
Ref.~\cite{xtcd-t47t}, we take $M_s=580$ kA/m, $A_1=2\times10^{-11}$ J/m,
$A_2=1\times10^{-11}$ J/m, $A_0=2\times10^{5}$ J/m$^3$,
$D=2.5\times10^{-3}$ J/m$^2$, $K=6\times10^{5}$ J/m$^3$, and film thickness $t=1$ nm.
These parameters give $\bar A=\sqrt{2}\times10^{-11}\,\mathrm{J/m}\simeq1.41\times10^{-11}\,\mathrm{J/m}$ and the reference wall-width scale $\lambda=\sqrt{\bar A/K}\simeq4.85\,\mathrm{nm}$.
We use $\rho=A_1/A_2$ for the exchange contrast and vary it at fixed $\bar A$.
\begin{figure*}[t]
\centering
\includegraphics[width=\textwidth]{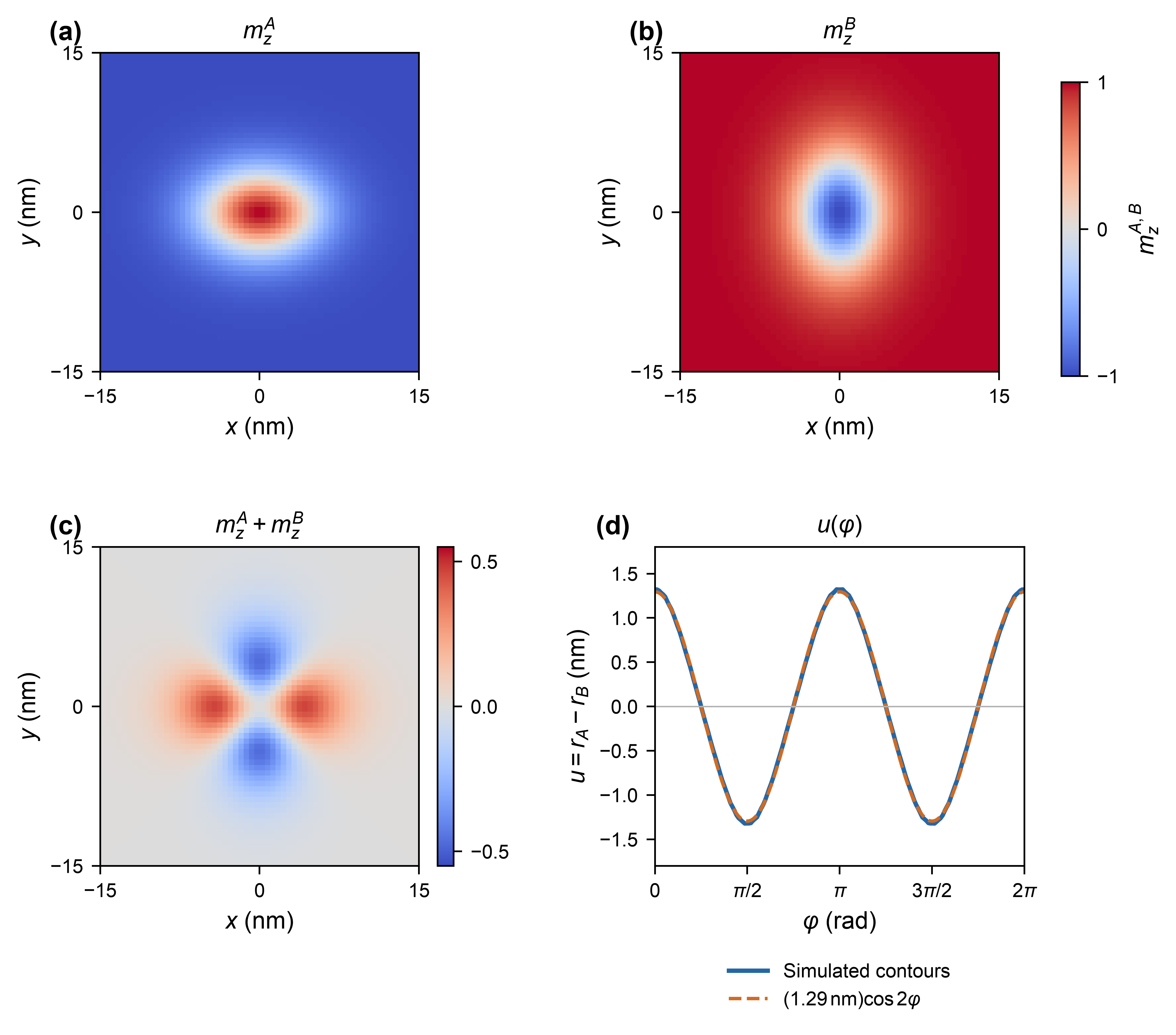}
\caption{Simulated $d$-wave altermagnetic skyrmion at zero field and $\rho=2$.
(a,b) Out-of-plane magnetization on sublattices $A$ and $B$.
(c) $m_z^A+m_z^B$.
(d) $u(\varphi)=r_A(\varphi)-r_B(\varphi)$, with $\varphi$ measured from the $x$ axis. The solid curve uses the simulated contours.
The dashed curve is $(1.29\,\mathrm{nm})\cos2\varphi$. }
\label{fig:quad}
\end{figure*}

Figures~\ref{fig:quad}(a) and \ref{fig:quad}(b) show the out-of-plane magnetization of the
two sublattices at zero field and $\rho=2$. The two $m_z=0$ contours are elliptical, with perpendicular long axes. 
The two radii are equal,
$R_A=R_B=3.87$ nm, and the aspect ratio is $b_A/a_A=1.40$.
Panel (c) shows the sum $m_z^A+m_z^B$. It forms four lobes of alternating sign.
The boundary of each sublattice component is its $m_{z}=0$ contour.
Where the $A$ boundary lies outside the $B$ boundary the local magnetization is positive, and
where the reverse holds it is negative, which is the alternating pattern of panel (c).
A compensated altermagnetic skyrmion therefore carries a spatially structured magnetization at
zero field even though its net moment vanishes.
In polar coordinates
centered on the skyrmion, $r_\eta(\varphi)$ is the distance from the center to that contour in
the direction $\varphi$, measured from the $x$ axis.
Panel (d) shows the angular variation of the mismatch between the two boundaries. The solid line gives \(u(\varphi)=r_A(\varphi)-r_B(\varphi)\), obtained from the simulated \(m_z=0\) contours. The dashed line shows \((1.29\,\mathrm{nm})\cos2\varphi\), with the amplitude extracted from these contours. The numerical $u(\varphi)$ is well approximated by $(1.29\,\mathrm{nm})\cos2\varphi$.The $\cos2\varphi$ dependence shows that the leading contour mismatch changes sign every $90^\circ$, as expected for
the two perpendicular elliptical contours. This angular mismatch
produces the four-lobed local magnetization in Fig.~\ref{fig:quad}(c).

\begin{figure*}[t]
\centering
\includegraphics[width=0.86\textwidth]{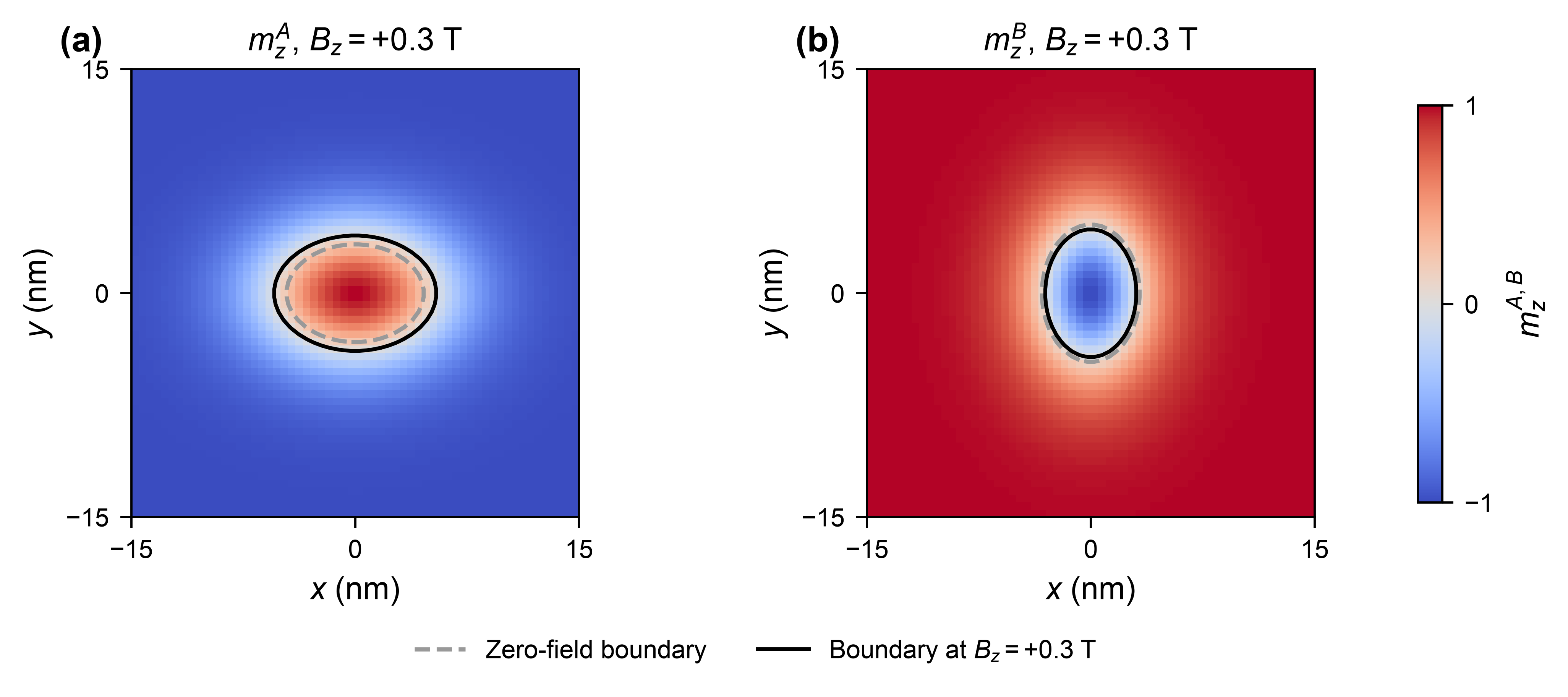}
\caption{Sublattice skyrmions at $B_z=+0.3$ T. The gray dashed curves show the $m_z=0$ contours at $B_z=0$, and the black curves show those at $B_z=+0.3\,\mathrm T$. (a) Sublattice $A$
expands to $R_A=4.58$ nm. (b) Sublattice $B$ contracts to $R_B=3.61$ nm.}
\label{fig:fieldstates}
\end{figure*}

At $B_z=+0.3$ T, Fig.~\ref{fig:fieldstates} shows that the $A$ skyrmion expands
from $3.87$ to $4.58$ nm and the $B$ skyrmion contracts to $3.61$ nm.
The radius difference is $\dR=0.97$ nm. Their mean radius increases to
$4.09$ nm, $5.67\%$ above its zero-field value.

\section{Magnetic field and radius difference}
\label{sec:coordinates}
\subsection{Coordinates and changes with the field}
At $B_z=0$, the net moment $\mu_z=t\int d^2r\,M_z$ vanishes even though $M_z$ is not zero everywhere. A small $\dR$ produces a contribution to $\mu_z$ proportional to $\dR$, so the Zeeman energy contains a term proportional to $-B_z\dR$. We describe the changes in a small field using the two radii and the two sublattice ellipticities.

We denote
the boundary distances along $x$ and $y$ by
\[
b_\eta=r_\eta(0),\qquad
a_\eta=r_\eta(\pi/2),
\]
and define
\[
\varepsilon_\eta=\frac{b_\eta-a_\eta}{b_\eta+a_\eta}.
\]
Positive $\varepsilon_\eta$ denotes elongation
along $x$, and negative
$\varepsilon_\eta$ denotes elongation along $y$. 
We use the following combinations of the two radii and ellipticities:
\begin{equation}
\begin{aligned}
\Rbar&=\tfrac12(R_A+R_B), &\quad \dR&=R_A-R_B,\\
\Qs&=\tfrac12(\varepsilon_A-\varepsilon_B), &\quad
\Nn&=\tfrac12(\varepsilon_A+\varepsilon_B).
\end{aligned}
\label{eq:multipoles}
\end{equation}
The definitions give
\[
R_A=\Rbar+\dR/2,\qquad R_B=\Rbar-\dR/2,
\]
so changing $\Rbar$ at fixed $\dR$ changes both radii equally, whereas changing $\dR$ at fixed $\Rbar$ changes them in opposite directions. Similarly,
\[
\varepsilon_A=\Qs+\Nn,\qquad \varepsilon_B=-\Qs+\Nn.
\]
At zero field, $\Nn=0$ and $\varepsilon_A=-\varepsilon_B$. At fixed $\Qs$, a change $\Delta \Nn$ gives $\Delta\varepsilon_A=\Delta\varepsilon_B=\Delta \Nn$.

Reflection about $x=y$, combined with sublattice exchange,
acts as
\begin{equation}
\begin{aligned}
\mathcal{P}:\quad
\mathbf{m}_A(x,y)&\to-\mathcal{R}\,\mathbf{m}_B(y,x),\\
\mathbf{m}_B(x,y)&\to-\mathcal{R}\,\mathbf{m}_A(y,x),
\end{aligned}
\label{eq:P}
\end{equation}
Here $\mathcal R(m_x,m_y,m_z)=(m_y,m_x,m_z)$. At zero field, all energy terms, including magnetostatics, are unchanged under $\mathcal P$. At fixed nonzero $B_z$, $\mathcal P$ reverses the Zeeman term. If $B_z$ is reversed simultaneously with $\mathcal P$, the Zeeman term is restored and the full energy is unchanged. The term-by-term verification is given in Supplemental Material, Sec.~S3.

The reflection sends $\varphi$ to $\pi/2-\varphi$. Together with sublattice exchange, it gives $R_A\leftrightarrow R_B$ and $\varepsilon_A\leftrightarrow-\varepsilon_B$. Thus,
\begin{equation}
\begin{aligned}
(\Rbar,\Qs)&\to(\Rbar,\Qs), && \mathcal{P}\text{-even},\\
(\dR,\Nn,B_z)&\to(-\dR,-\Nn,-B_z), && \mathcal{P}\text{-odd}.
\end{aligned}
\label{eq:grading}
\end{equation}

For fixed values of $(\Rbar,\Qs,\dR,\Nn)$ and $B_z$, we define $E(\Rbar,\Qs,\dR,\Nn;B_z)$ as the minimum of the full spin energy over configurations with these collective coordinates. Eq.~(\ref{eq:grading}) makes this energy invariant under simultaneous reversal of $(\dR,\Nn,B_z)$. At quadratic order, $\mathcal P$ permits $B_z\dR$, $B_z\Nn$, and $\dR \Nn$, but excludes terms linear in $\dR$, $\Nn$, or $B_z$ individually. We therefore write
\begin{equation}
\begin{aligned}
E={}&E_0(\Rbar,\Qs;B_z^2)
-B_z(\Lambda_R\dR+\Lambda_\Nn\Nn)\\
&+\frac{1}{2}\kappa_0\dR^2
+\frac{1}{2}\kappa_\Nn\Nn^2+\zeta\,\dR\,\Nn+\mathcal{O}(4).
\end{aligned}
\label{eq:Eexp}
\end{equation}
At fixed $\Rbar$ and $\Qs$, the coefficients are evaluated at $B_z=\dR=\Nn=0$. The coefficient $\kappa_0$ is the second derivative of $E$ with respect to $\dR$ while $\Nn=0$, and $\kappa_{\Nn}$ is the second derivative with respect to $\Nn$ while $\dR=0$. They determine the quadratic energy costs of the two deformations. The term $\zeta\dR\Nn$ couples the radius difference to the uniform ellipticity. The coefficients $\Lambda_R$ and $\Lambda_{\Nn}$ give the terms proportional to $\dR$ and $\Nn$ in the net moment at zero field. The term $E_0$ is independent of these two coordinates and is even in $B_z$. Here $O(4)$ denotes fourth and higher total order in $(\dR,\Nn,B_z)$, with $\Rbar$ and $\Qs$ held fixed.

A small $B_z$
changes $\dR$ and $\Nn$ linearly, whereas the changes in $\Rbar$
and $\Qs$ begin at order $B_z^2$:
\[
\Rbar(B_z)=\Rbar(0)+O(B_z^2),\qquad
\Qs(B_z)=\Qs(0)+O(B_z^2).
\]
Minimizing the quadratic energy in Eq.~(\ref{eq:Eexp}) with respect to $\Nn$ and $\dR$ gives, to first order in $B_z$,
\begin{equation}
\begin{aligned}
\Nn&=\frac{\Lambda_\Nn B_z-\zeta\dR}{\kappa_\Nn},\\
\dR&=\frac{\Lambda_R-\zeta\Lambda_\Nn/\kappa_\Nn}{\kappa}B_z
\equiv\chi B_z,\\
\kappa&=\kappa_0-\frac{\zeta^2}{\kappa_\Nn}.
\end{aligned}
\label{eq:chi}
\end{equation}
The first equation contains a term $\Lambda_\Nn B_z/\kappa_\Nn$ from the field and a term $-\zeta\dR/\kappa_\Nn$ from the coupling to $\dR$.
To determine the zero-field energy cost of a small $\dR$, we fix $\Rbar$, $\Qs$, and $\dR$ and minimize over $\Nn$. For $\kappa_{\Nn}>0$, the quadratic energy gives
\[
\Nn=-\frac{\zeta}{\kappa_{\Nn}}\dR .
\]
After this minimization, the energy increase relative to the
$\dR=0$ state is
\[
\Delta E(\dR)
=
\frac12\kappa\dR^2+O(\dR^4),
\qquad
\kappa=\kappa_0-\frac{\zeta^2}{\kappa_{\Nn}}.
\]
For $\kappa_\Nn>0$, allowing $\Nn$ to change lowers the energy by $\zeta^2\dR^2/(2\kappa_\Nn)$ relative to holding $\Nn=0$. Thus $\kappa\leq\kappa_0$, with equality when $\zeta=0$. The quadratic energy in $(\dR,\Nn)$ is positive when $\kappa_\Nn>0$ and $\kappa>0$. At $\kappa=0$, higher-order terms are needed to determine stability.
\begin{figure*}[t]
\centering
\includegraphics[width=0.95\textwidth]{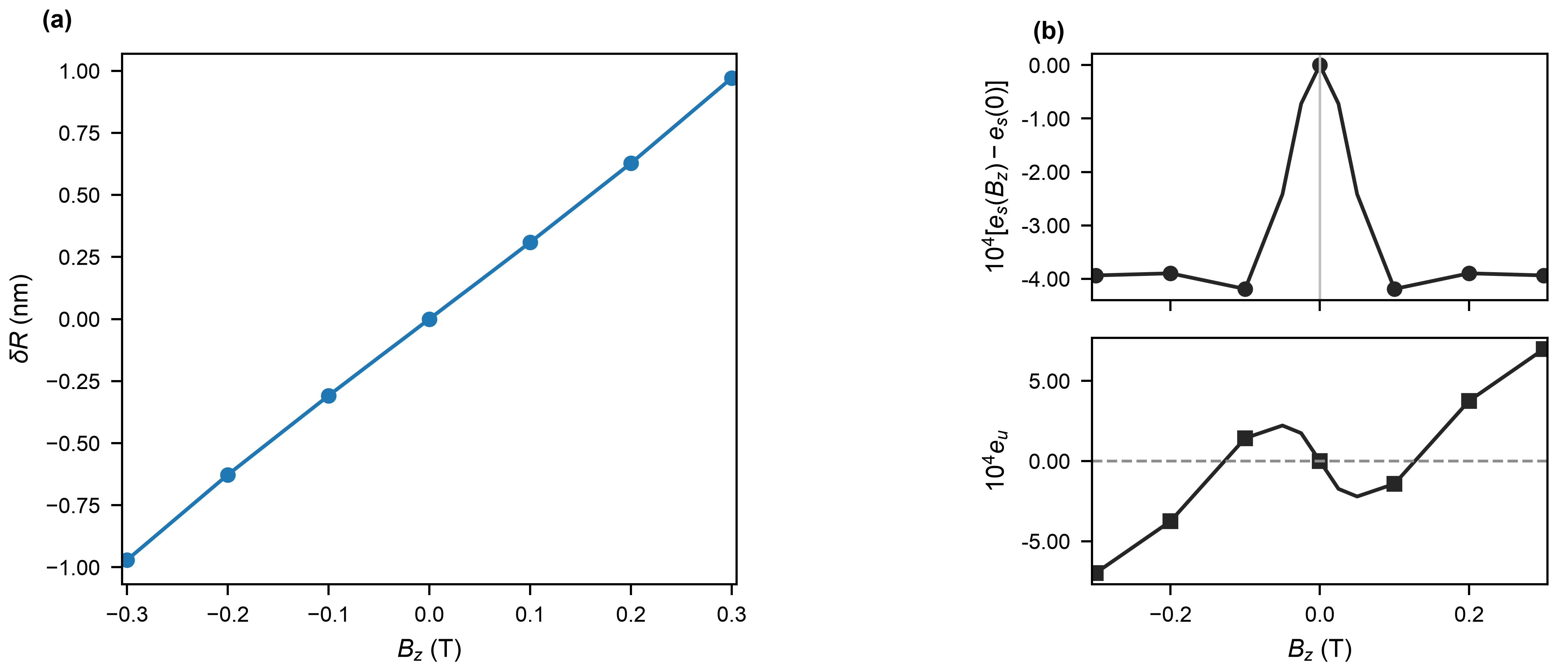}
\caption{Changes in the radii and ellipticities at $\rho=2$. (a) $\dR$ versus $B_z$. (b) $10^4[\Qs(B_z)-\Qs(0)]$ and $10^4\Nn(B_z)$, showing the even and odd dependences, respectively. }
\label{fig:field}
\end{figure*}
The slope $\chi=(d\dR/dB_z)_0$ measures the radius difference produced by a small field.
Using the states at $B_z=\pm0.05$ T, we obtain
\begin{equation}
\chi\simeq\frac{\dR(0.05\,\mathrm T)-\dR(-0.05\,\mathrm T)}{0.10\,\mathrm T}
=3.08\ \mathrm{nm/T}.
\label{eq:chimeas}
\end{equation}
The symmetric differences are described in Supplemental Material, Sec.~S4.
Figure~\ref{fig:field}(a) shows the increase of $\dR$ with $B_z$. In Fig.~\ref{fig:field}(b), reversing $B_z$ leaves $\Qs$ unchanged and reverses $\Nn$, consistent with Eq.~(\ref{eq:grading}). Over the plotted interval, $|\Nn|<10^{-3}$, compared with $\Qs(0)\simeq1.68\times10^{-1}$. A small positive $B_z$ therefore increases $R_A$ and decreases $R_B$ while changing the signed ellipticities only slightly. For the symmetric field pairs $B_z=\pm B_1$ with $B_1=0.025$ and $0.05\,\mathrm{T}$, the differences $[\Nn(B_1)-\Nn(-B_1)]/(2B_1)$ are negative. At larger positive fields, $\Nn$ changes sign. 
The additional calculations are reported in Supplemental Material, Sec.~S4.

In summary,$\dR$ is linear in $B_z$, while the
change in $\Rbar$ begins at order $B_z^2$. The uniform ellipticity
$\Nn$ can also change linearly with $B_z$, but it remains below
$10^{-3}$ over the plotted range, and $\Qs$ has no linear change.
Thus, the first-order size change produced by the field is the
radius difference $\dR$, rather than a common change of the two
radii. This is why we use $\dR$ as the radial coordinate for the
field-driven oscillation studied in Sec.~VI.

\subsection{From the contours to the magnetic moment}
We now relate the contour difference $u(\varphi)$ to the local magnetization and the net moment, explaining why a uniform $B_z$ couples to $\dR$. We approximate $m_{Az}$ and $-m_{Bz}$ by the same decreasing radial profile $f$, with $f(0)=+1$, $f(1)=0$, and $f(\infty)=-1$:
\begin{equation}
m_{Az}=f\!\left(\frac{r}{r_A(\varphi)}\right),\qquad
m_{Bz}=-f\!\left(\frac{r}{r_B(\varphi)}\right).
\label{eq:profile}
\end{equation}
To first order in ellipticity, the contour is
\begin{equation}
r_\eta(\varphi)\simeq R_\eta[1+\varepsilon_\eta\cos2\varphi].
\label{eq:contour}
\end{equation}
The expansion through second order is derived in Supplemental Material, Sec.~S5. Subtracting the two contours gives
\begin{equation}
\begin{aligned}
u(\varphi)&=r_A(\varphi)-r_B(\varphi)\\
&\simeq\dR+\big(2\Rbar\Qs+\dR\Nn\big)\cos2\varphi.
\end{aligned}
\label{eq:u}
\end{equation}

Expanding $f(r/r_A)$ and $f(r/r_B)$ to first order about $r_A=r_B=\Rbar$ gives
\begin{equation}
\begin{aligned}
M_z(r,\varphi)&=M_s(m_{Az}+m_{Bz})\simeq M_sp(r)u(\varphi),\\
p(r)&=-\frac{r}{\Rbar^2}f'\!\left(\frac{r}{\Rbar}\right)>0.
\end{aligned}
\label{eq:Mz}
\end{equation}
The radial factor $p(r)$ is concentrated near the wall, while $u(\varphi)$ determines the angular sign of $M_z$. The derivation is given in Supplemental Material, Sec.~S5.

At $B_z=0$, $u(\varphi)\simeq2\Rbar\Qs\cos2\varphi$, giving four lobes of $M_z$ with alternating signs and zero net moment \cite{PhysRevLett.133.196701,xtcd-t47t}. At $\rho=2$, $u(\varphi)\simeq(1.29\,\mathrm{nm})\cos2\varphi$, with an amplitude of about $0.27\lambda$ for $\lambda=4.85\,\mathrm{nm}$. 
Angular integration gives
\begin{equation}
\mu_z=t\!\int d^2r\,M_z
=2\pi M_st\,\dR\!\int_0^\infty r p(r)\,dr,
\label{eq:mu}
\end{equation}
to first order, with $\Lambda_R=2\pi M_st\int_0^\infty rp(r)\,dr$. For a circular contour with $f(\xi)=+1$ for $\xi<1$ and
$f(\xi)=-1$ for $\xi>1$,
\begin{equation}
\Lambda_R=4\pi M_st\Rbar.
\label{eq:Lambda}
\end{equation}
For elliptical contours and the scaled profile in Eq.~(\ref{eq:profile}), each core integral is proportional to $R_\eta^2$, independently of $\varepsilon_\eta$. Consequently, $\mu_z\propto R_A^2-R_B^2=2\Rbar\dR$. The higher magnetic moments of the out-of-plane and in-plane magnetization are derived in Supplemental Material, Sec.~S6. 

A uniform $B_z$ couples to the integral of $M_z$, giving
\begin{equation}
E_Z=-B_z\,t\!\int\!d^2r\,M_z=-B_z\mu_z=-\Lambda_R B_z\,\dR.
\label{eq:EZ}
\end{equation}
The $\cos2\varphi$ term integrates to zero. 
Changing $\Nn$ at fixed $R_A$ and $R_B$ leaves $\mu_z$ unchanged, giving $\Lambda_{\Nn}=0$. The calculations in Supplemental Material, Sec.~S7 give a nonzero estimate of $\Lambda_{\Nn}$ for the numerical profiles. 
Since $\Lambda_R>0$ for the chosen core polarities,
a positive $B_z$ favors a positive $\dR$. 

Hence,
the altermagnetic skyrmion can have a nonzero local magnetization while its net moment vanishes.
$\cos2\varphi$ part of $u(\varphi)$ produces the four-lobed zero-field pattern, whereas the angle-independent part $\dR$ produces a net moment to first order.

\section{Radius difference induced by exchange modulation}
\label{sec:knobs}
\label{sec:strain}

To study a radius difference generated without a magnetic field, we model strain by applying the same directional exchange modulation to both sublattices at $B_z=0$:
\begin{equation}
\begin{aligned}
(A_{Ax},A_{Ay})&=\big(A_1(1{+}\delta),\,A_2(1{-}\delta)\big),\\
(A_{Bx},A_{By})&=\big(A_2(1{+}\delta),\,A_1(1{-}\delta)\big),
\end{aligned}
\label{eq:strainA}
\end{equation}
Here $\delta$ is the dimensionless relative change of the exchange coefficients: the coefficients along $x$ change by $+\delta$ and those along $y$ by $-\delta$, relative to their unperturbed values. Its relation to lattice strain requires the material dependence of the exchange coefficients. The anisotropy and DMI are held fixed. The factors are the same on both sublattices, but the absolute changes differ when $A_1\ne A_2$. For both sublattices, the geometric mean is $\bar A\sqrt{1-\delta^2}=\bar A[1-\delta^2/2+O(\delta^4)]$.  For a fixed dimensionless N\'eel profile and aspect ratio $b_\eta/a_\eta$, minimization of the local single-sublattice energy with respect to $R_\eta$ gives
\begin{equation}
R_\eta=R_0\left(\sqrt{\frac{b_\eta}{a_\eta}}+
\sqrt{\frac{a_\eta}{b_\eta}}\right),
\label{eq:Rg}
\end{equation}
where $R_0\propto D/K$ depends on the assumed profile. This estimate omits intersublattice exchange and magnetostatics. Its derivation is given in Supplemental Material, Sec.~S8.

Minimizing only the exchange energy gives $b_A/a_A=\sqrt\rho$ and $b_B/a_B=1/\sqrt\rho$. The modulation multiplies both ratios by $[(1+\delta)/(1-\delta)]^{1/2}$.
Equation~(\ref{eq:Rg}) gives an increase of $R_A$ and a decrease of $R_B$. To first order,
\begin{equation}
\dR\simeq R_0\big(\rho^{1/4}-\rho^{-1/4}\big)\delta,
\label{eq:strain}
\end{equation}
for the assumed local profile. We define $\chi_\epsilon=(d\dR/d\delta)_0$ as the slope of the radius difference with respect to the exchange modulation. We obtain it by minimizing the full energy at each $\delta$ and fitting $\dR(\delta)$ over $|\delta|\leq0.02$. The derivation and numerical values are given in Supplemental Material, Sec.~S8.

\begin{figure*}[t]
\centering
\includegraphics[width=0.95\textwidth]{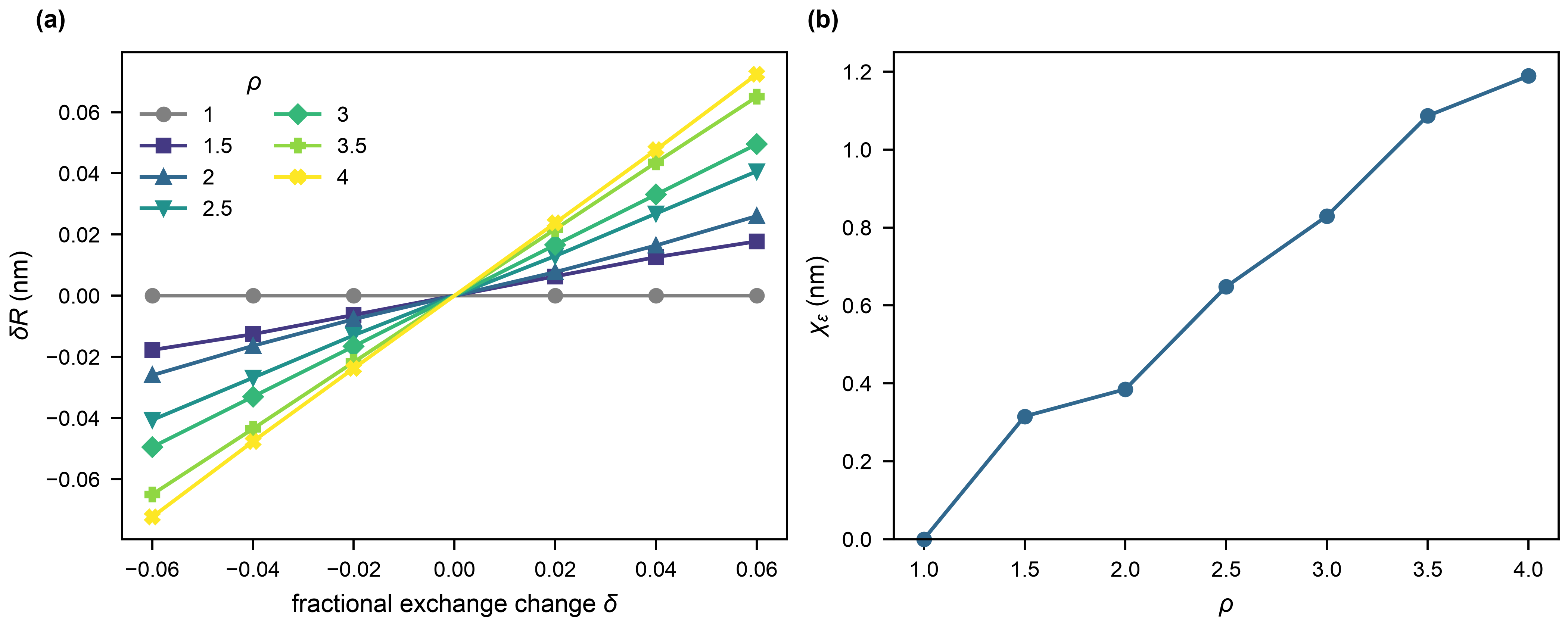}
\caption{Radius difference produced by the exchange modulation at $B_z=0$. All coefficients along $x$ are multiplied by $1+\delta$ and those along $y$ by $1-\delta$ on both sublattices. The unperturbed $\bar A=\sqrt{A_1A_2}$ and the other parameters are fixed as in Sec.~\ref{sec:model}. (a) $\dR$ versus $\delta$ for the listed $\rho=A_1/A_2$. Reversing $\delta$ reverses $\dR$, which vanishes at $\rho=1$. (b) $\chi_\epsilon=(d\dR/d\delta)_0$ versus $\rho$.}
\label{fig:strain}
\end{figure*}

Figure~\ref{fig:strain}(a) shows $\dR(\delta)$ for $-0.06\leq\delta\leq0.06$. 
Figure~\ref{fig:strain}(b) shows $\chi_\epsilon$ as a function of $\rho$. It is $3.85\times10^{-1}$ nm at $\rho=2$ and $1.19$ nm at $\rho=4$. At $\rho=1$, the two sublattices have identical exchange tensors and $\chi_\epsilon=0$, consistent with the numerical result. As $|\delta|$ increases, the curves in Fig.~\ref{fig:strain}(a) depart from the linear relation $\dR=\chi_\epsilon\delta$.

The interchanged exchange directions therefore allow the same directional modulation on both sublattices to change their radii oppositely at zero field. For the modulation studied here, $\chi_\epsilon$ vanishes when the sublattice exchange tensors are identical and increases over the sampled $\rho>1$. This gives a way to change the internal radius difference through the exchange coefficients, distinct from the magnetic-field coupling, which remains nonzero at $\rho=1$.

\section{Equilibrium size and radius difference}
\label{sec:stiffness}
We test how the relative orientation of the two sublattice textures affects the size and field-induced radius difference of an altermagnetic skyrmion. We denote the arrangement with $(A_1,A_2)$ along $(x,y)$ on sublattice A and $(A_2,A_1)$ on sublattice B by AM. Its two sublattice skyrmions have perpendicular long axes. In the reference PAR, both sublattices have $(A_1,A_2)$ along $(x,y)$, and the long axes are parallel. We compare their zero-field mean radii $\Rbar$ and small-field slopes $\chi=(d\dR/dB_z)_0$. For AM at $\rho=2$, we then calculate the zero-field energy of unequal-radius states to construct the radial part of the oscillation model.

\subsection{Perpendicular and parallel exchange axes}
All coefficients other than the assignment of $A_1$ and $A_2$ to the two sublattices are unchanged, as are the film thickness and magnetostatic calculation. PAR is a collinear antiferromagnetic reference with anisotropic in-plane exchange. The subscript $\parallel$ denotes PAR. At $\rho=1$, the two models coincide.

\begin{figure*}[t]
\centering
\includegraphics[width=0.95\textwidth]{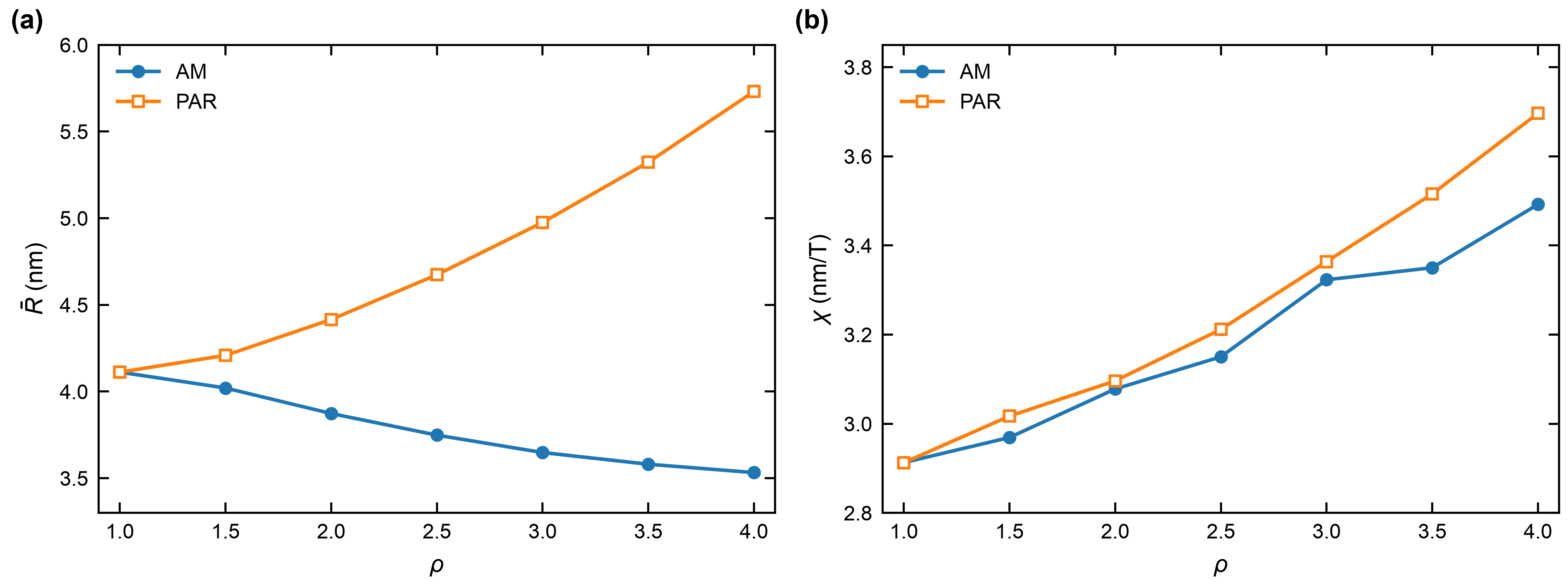}
\caption{AM and PAR compared at fixed $\bar A$ versus $\rho=A_1/A_2$. (a) Mean radius $\Rbar$ at $B_z=0$. (b) $\chi=(d\dR/dB_z)_0$ using the symmetric difference
$\chi \simeq
\frac{\dR(B_1)-\dR(-B_1)}{2B_1},
B_1=0.05~\mathrm{T}.$
Both geometries use the same $96$ nm cell, $h=0.5$ nm grid, and remaining parameters. Blue filled circles denote AM and orange open squares denote PAR. }
\label{fig:control}
\end{figure*}

Figure~\ref{fig:control}(a) and Table~\ref{tab:scan} compare the zero-field radii. AM has the smaller $\Rbar$ at every sampled $\rho>1$. At $\rho=4$, $\Rbar_{\rm AM}=3.53$ nm and $\Rbar_{\parallel}=5.73$ nm.

Figure~\ref{fig:control}(b) compares estimates of $\chi$ obtained from the states at $B_z=\pm0.05\,\mathrm{T}$ on the $h=0.5\,\mathrm{nm}$ grid. At $\rho=4$, $\chi$ is $5.53\%$ smaller in AM than in PAR. The moment change per unit radius difference, $\Lambda_{\rm path}=\Delta\mu_z/\Delta\dR$, is $28.36\%$ smaller. Here $\Delta X=X(0.05\,\mathrm{T})-X(-0.05\,\mathrm{T})$, and both percentages use PAR as the reference. Equation~(\ref{eq:chi}) shows that $\chi$ depends on both the magnetic coupling and $\kappa$. The states and differences are specified in Supplemental Material, Sec.~S4.

At $\rho=2$, reducing $h$ from $0.50$ to $0.25$ nm in the same $96$ nm cell changes $\chi_{\parallel}-\chi_{\mathrm{AM}}$ from $1.83\times10^{-2}$ to $6.17\times10^{-2}$ nm/T, using $B_z=\pm0.05$ T. PAR has the larger value on both grids, while the magnitude of the difference remains sensitive to $h$.

\begin{table*}[t]
\caption{Zero-field mean radii and small-field slopes for AM and PAR at fixed $\bar A=\sqrt{2}\times10^{-11}$ J/m. The radii use $R_\eta=\sqrt{a_\eta b_\eta}$. Each $\chi$ is estimated as $[\dR(B_1)-\dR(-B_1)]/(2B_1)$, with $B_1=0.05$ T. The entries use $h=0.5$ nm in a $96$ nm periodic cell. The additional grid comparison at $\rho=2$ is reported in Sec.~\ref{sec:stiffness}\,A. AM and PAR coincide at $\rho=1$.}
\label{tab:scan}
\begin{ruledtabular}
\begin{tabular}{c c c c c}
$\rho$ & $\Rbar_{\rm AM}$ (nm) & $\Rbar_{\parallel}$ (nm) & $\chi_{\rm AM}$ (nm/T) & $\chi_{\parallel}$ (nm/T)\\
\colrule
1 & 4.11 & 4.11 & 2.91 & 2.91\\
1.5 & 4.02 & 4.21 & 2.97 & 3.02\\
2 & 3.87 & 4.42 & 3.08 & 3.10\\
2.5 & 3.75 & 4.67 & 3.15 & 3.21\\
3 & 3.65 & 4.98 & 3.32 & 3.36\\
3.5 & 3.58 & 5.32 & 3.35 & 3.52\\
4 & 3.53 & 5.73 & 3.49 & 3.70\\
\end{tabular}
\end{ruledtabular}
\end{table*}

The local wall model in Supplemental Material, Sec.~S9 identifies an energy cost of the noncoincident contours. Here $E_{\rm inter}=A_0t\int d^2r\,[1+\mathbf m_A\cdot\mathbf m_B]$ is the excess intersublattice energy relative to exactly antiparallel spins. For two walls of the same width $\lambda$, the model gives
\[
E_{\rm inter}\simeq A_0t\lambda\oint \Rbar\,d\varphi\,
F\!\left(\frac{u(\varphi)}{\lambda}\right),
\]
where $F$ is even, vanishes at zero, and increases with the magnitude of its argument for the profiles used there. In AM, $u(\varphi)\simeq2\Rbar\Qs\cos2\varphi$ already at $\dR=0$. Increasing $\Rbar$ while keeping the shape and $\lambda$ fixed raises this intersublattice energy. In the symmetric PAR state, $u=0$ and the two magnetizations cancel at every point, so this energy is zero. The equilibrium radii in Fig.~\ref{fig:control}(a) are obtained by minimizing all terms in Eq.~(\ref{eq:w}).

AM also has a nonzero magnetostatic energy. At $\rho=2$, the contributions from magnetization in the film plane and along $z$ are $8.96\times10^{-22}$ J and $2.18\times10^{-21}$ J, respectively. Both are included in the numerical minimization.

Across $\rho=1$ to $4$, $\Rbar_{\rm AM}$ decreases by about $14.09\%$, whereas $\Rbar_{\parallel}$ increases by about $39.37\%$. Over the same scan, the small-field slopes increase in both geometries. Thus the AM skyrmion becomes smaller as $\rho$ increases, while a given small field produces a larger radius difference.

These results show that the size of an altermagnetic skyrmion is controlled not only by the properties of each sublattice texture, but also by their relative geometry. 

\subsection{Energy of unequal radii}
\label{sec:radial_family}

To construct the radial part of the oscillation model, we calculate a family of states at $B_z=0$ and $\rho=2$ in AM. For each prescribed
\[
q=\sqrt{S_A/\pi}-\sqrt{S_B/\pi},\qquad
S_\eta=\frac12\int d^2r\,[1+s_\eta m_{\eta z}],
\]
with $s_A=+1$ and $s_B=-1$, we minimize the energy over the remaining spin variables. We then measure $\dR=\sqrt{a_A b_A}-\sqrt{a_B b_B}$ from the resulting contours.

\begin{figure}[t]
\centering
\includegraphics[width=0.95\columnwidth]{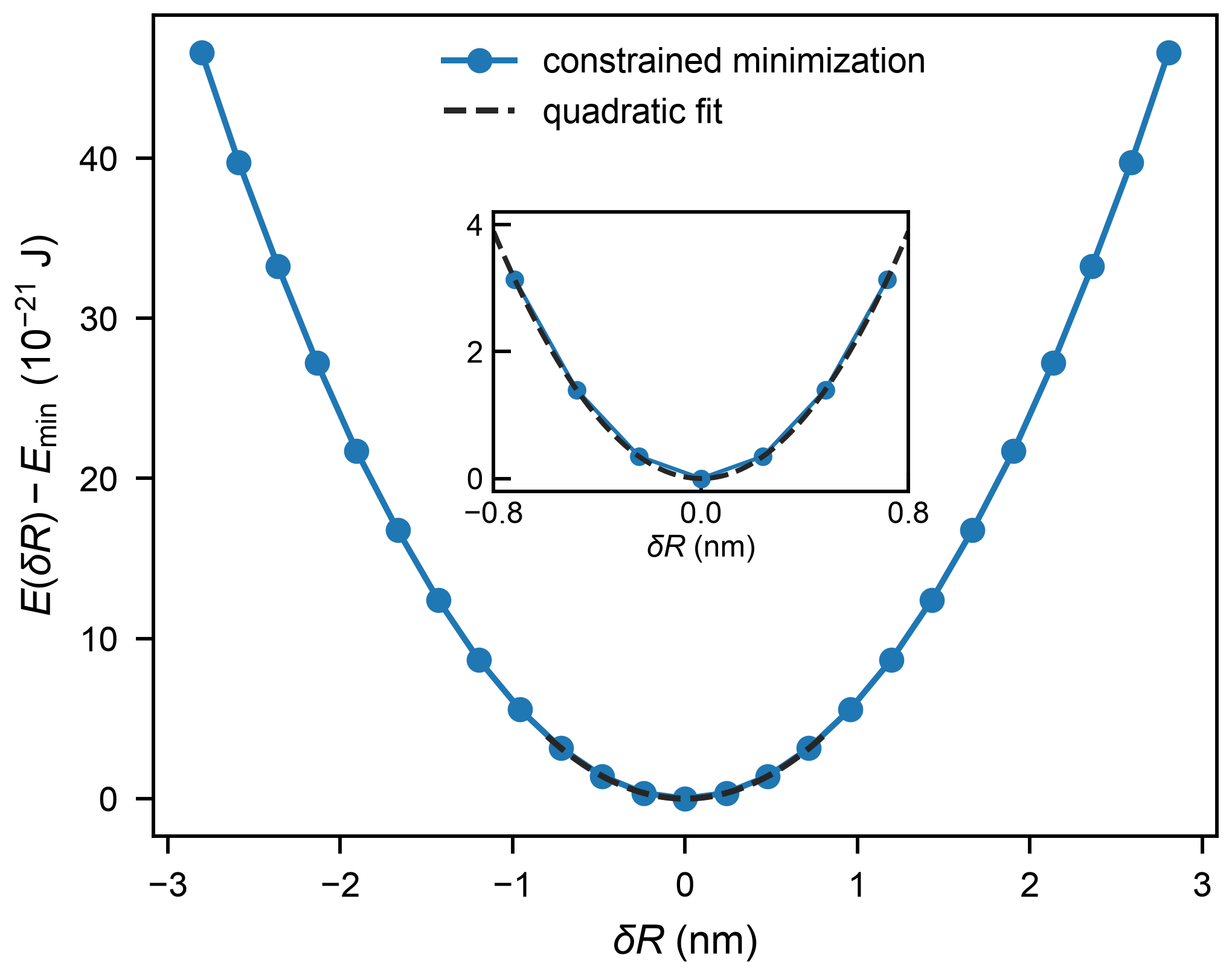}
\caption{Energy of AM states minimized at fixed $q$, plotted against the measured $\dR$, at $B_z=0$ and $\rho=2$. Energies are measured relative to the zero-field state at $\dR=0$. All plotted states retain both cores. The dashed curve is a fit of $E=E_c+\tfrac12\kappa_{\rm con}\dR^2$ over $|\dR|\leq0.8$ nm. The inset shows the interval used in the fit.}
\label{fig:well}
\end{figure}

The states in Fig.~\ref{fig:well} retain both cores for $|\dR|\lesssim2.805$ nm. Their energy is lowest at $\dR=0$. A fit of $E=E_c+\tfrac12\kappa_{\rm con}\dR^2$ over $|\dR|\leq0.8\,\mathrm{nm}$ gives $\kappa_{\rm con}=12.17\,\mathrm{mJ/m^2}$. The intercept $E_c$ is fitted independently. In Sec.~\ref{sec:coordinates}\,A, $\kappa$ is defined by minimizing the energy over $\Nn$ at fixed $\dR$. Here $\kappa_{\rm con}$ is fitted to states minimized at fixed $q$. We use this family and $\kappa_{\rm con}$ for the radial part of the model in Sec.~\ref{sec:dynamics}. The constraints and fit are described in Supplemental Material, Sec.~S10.

These states identify an internal deformation of the altermagnetic skyrmion in which the two sublattice radii become unequal while both cores remain present. Along the selected fixed-$q$ family, a small $\dR$ raises the zero-field energy by $\tfrac12\kappa_{\rm con}\dR^2$. This energy cost supplies the radial restoring term in the oscillation model of Sec.~VI.

\section{Coupled oscillations of radius difference and helicity}
\label{sec:dynamics}

We apply a short perpendicular field pulse and record $\dR(t)$ after the field is removed. We also measure the helicity of each sublattice relative to the radial direction. The two equilibrium helicities differ by $\pi$. After subtracting each sublattice's zero-field value, we denote the helicity deviations by $\phi_A$ and $\phi_B$. We define the uniform and staggered deviations as $\bar\phi=(\phi_A+\phi_B)/2$ and $\phi_-=(\phi_A-\phi_B)/2$. Their extraction from the spins and the time-integration and fitting procedures are described in Supplemental Material, Sec.~S11.

\begin{figure*}[t]
\centering
\includegraphics[width=0.95\textwidth]{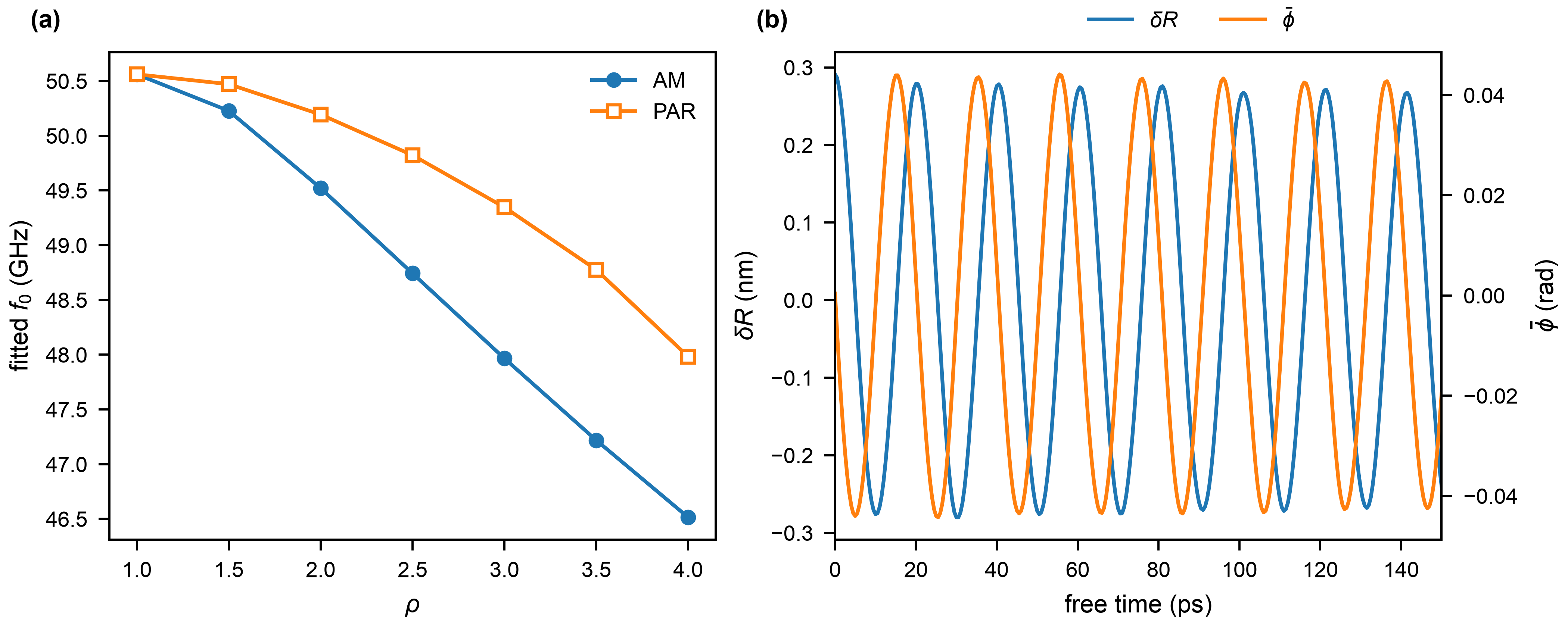}
\caption{Oscillations after a perpendicular field pulse. (a) Fitted frequency versus $\rho=A_1/A_2$ for AM and PAR. (b) $\dR(t)$ on the left axis and $\bar\phi(t)$ in radians on the right axis for AM at $\rho=2$. The pulse has amplitude $0.05$ T and duration $10$ ps. The subsequent calculation uses $B_z=0$ and $\alpha=10^{-3}$. Both $\dR(t)$ and $\bar\phi(t)$ in (b) are measured from their
zero-field equilibrium values. }
\label{fig:dyn}
\end{figure*}

We apply $B_z=0.05$ T for $10$ ps, set $B_z=0$, and integrate the LLG equations for $2$ ns with $\alpha=10^{-3}$. At $\rho=2$, the fitted frequency is $49.52$ GHz and the ratio of the fitted oscillation amplitudes is $|\bar\phi|/|\dR|=1.57\times10^8$ rad/m. With the cosine convention used in the fit, $\bar\phi$ leads $\dR$ by $90.01^\circ$, approximately one quarter of a period.
Figure~\ref{fig:dyn}(b) shows the phase difference between $\dR$ and $\bar\phi$. Figure~\ref{fig:dyn}(a) compares the fitted frequencies in AM and PAR as $\rho$ is varied. With the other parameters fixed, the AM frequency decreases from $50.56$ GHz at $\rho=1$ to $46.51$ GHz at $\rho=4$, where the PAR value is $47.98$ GHz.

Under $\mathcal P$, $\bar\phi$ changes sign and $\phi_-$ does not. Among the retained coordinates, only $(\dR,\Nn,\bar\phi)$ can be excited to first order in the amplitude of a uniform $B_z$ pulse. The oscillating parts of $\Rbar$, $\Qs$, and $\phi_-$ vanish at this order. At the $\dR$ peak, the Fourier amplitudes of $\Rbar$ and $\phi_-$ are much smaller than those of $\dR$ and $\bar\phi$, respectively. These amplitudes, together with that of $\Nn$, are reported in Supplemental Material, Sec.~S11.

We approximate the motion by the family of states minimized at fixed $q$ in Sec.~\ref{sec:radial_family}, labelled by the measured $\dR$, together with a common additional spin rotation $\bar\phi$ about $z$. This defines a variational model with coordinates $\dR$ and $\bar\phi$. We evaluate its Berry contribution as follows. Define $I_\eta=\int d^2r\,(1+s_\eta\cos\theta_\eta)$, where $\theta_\eta$ is the polar angle of $\mathbf m_\eta$, $s_A=+1$, and $s_B=-1$. For spatially uniform increments $\phi_A$ and $\phi_B$ of the spin azimuth relative to the static profiles, the Berry contribution is
\begin{equation}
S_{\rm B}=\frac{M_st}{\gamma}\int\! dt\,
\Big[-I_A\dot\phi_A+I_B\dot\phi_B\Big].
\label{eq:SB}
\end{equation}
The integrands in $I_A$ and $I_B$ vanish in the respective backgrounds. The sign convention is derived in Supplemental Material, Sec.~S12. Substituting $\phi_{A,B}=\bar\phi\pm\phi_-$ gives
\begin{equation}
S_{\rm B}=-\frac{M_st}{\gamma}\int\! dt\,
\Big[\big(I_A-I_B\big)\dot{\bar\phi}
    +\big(I_A+I_B\big)\dot\phi_-\Big].
\label{eq:SB2}
\end{equation}
The definitions give $M_st(I_A-I_B)=\mu_z$, so $I_A-I_B=0$ in the symmetric state. Its first-order change along the fixed-$q$ family gives the Berry term $-P\dR\dot{\bar\phi}$. The coefficient $P$, evaluated at $q=\dR=0$, is
\begin{equation}
P=\frac{M_st}{\gamma}
\left.\frac{d(I_A-I_B)}{d\dR}\right|_0
\label{eq:Pnum}
\end{equation}
Equivalently, $P=\gamma^{-1}d\mu_z/d\dR$, evaluated along the same states minimized at fixed $q$. The sum $I_A+I_B$ multiplies $\dot\phi_-$ and changes when both radii change together. The analogous pairing of radius and spin angle has been used for a single skyrmion and coupled layers \cite{PhysRevB.99.054430,PhysRevB.102.104403}.

The frequency also requires the energy cost of changing $\bar\phi$. We set $\phi_A=\phi_B=\bar\phi$ while keeping the contours and polar angle profiles fixed, and define this cost as $\tfrac12\kappa_{\bar\phi}\bar\phi^2$. This rotation leaves the local exchange, anisotropy, and intersublattice terms unchanged. It changes the DMI and magnetostatic energies. For two circular N\'eel walls with width negligible compared with $\Rbar$, the estimate is
\begin{equation}
\kappa_{\bar\phi}\simeq4\pi^2 D\Rbar t+\kappa_{\bar\phi}^{\rm ms},
\label{eq:kphi}
\end{equation}
We obtain $\kappa_{\bar\phi}$ from the second-order energy change under a common rotation of the spins in the zero-field state. At $\rho=2$, the fit gives $\kappa_{\bar\phi}=4.08\times10^{-19}\,\mathrm{J}$. The separate DMI and magnetostatic contributions are given in Supplemental Material, Sec.~S13.

The product $\dR\bar\phi$ is unchanged under $\mathcal P$. A mirror in $x$ combined with time reversal leaves $\dR$ unchanged and reverses $\bar\phi$, which forbids this term. The verification is given in Supplemental Material, Sec.~S14. For this trial family, the quadratic Lagrangian without damping is
\[
L_{\rm var}=-P\dR\dot{\bar\phi}-\frac12\kappa_{\rm con}\dR^2
-\frac12\kappa_{\bar\phi}\bar\phi^2.
\]
The equations are $P\dot{\dR}=\kappa_{\bar\phi}\bar\phi$ and $P\dot{\bar\phi}=-\kappa_{\rm con}\dR$. They give $m^*\ddot{\dR}=-\kappa_{\rm con}\dR$, with
\begin{equation}
m^*=\frac{P^2}{\kappa_{\bar\phi}},
\label{eq:mstar}
\end{equation}
and $\omega_0=\sqrt{\kappa_{\rm con}\kappa_{\bar\phi}}/|P|$. Here $m^*$ is the effective mass associated with the radius-difference coordinate $\dR$. The corresponding frequency is $f_0=\omega_0/(2\pi)$. The circular-wall estimate of $m^*$ for $\lambda/\Rbar\ll1$ is derived
in Supplemental Material, Sec.~S14. The role of internal spin deformations in skyrmion inertia is discussed in Refs.~\cite{PhysRevLett.109.217201,PhysRevB.90.174434,PhysRevB.111.144411}.

For AM at $\rho=2$, the states minimized at fixed $q$ give
$P=2.12\times10^{-22}\,\mathrm{J\,s/m}$ and
$\kappa_{\rm con}=12.17\,\mathrm{mJ/m^2}$.
Together with $\kappa_{\bar\phi}$ from the common spin rotation,
the unrounded coefficients give $f_0=53.00\,\mathrm{GHz}$ and
$|\bar\phi|/|\dR|=1.73\times10^8\,\mathrm{rad/m}$,
compared with $49.52\,\mathrm{GHz}$ and
$1.57\times10^8\,\mathrm{rad/m}$ from LLG.
The frequency and amplitude ratio are $7.03\%$ and $9.58\%$
above the LLG values, respectively, using the unrounded coefficients.
The same coefficients give $m^*=1.10\times10^{-25}\,\mathrm{kg}$.
The dependence of $\Nn$ on the chosen family and its small oscillation
in LLG are examined in Supplemental Material, Sec.~S15.

The altermagnetic skyrmion therefore supports a field-excitable internal mode in which its two sublattice radii oscillate oppositely and the uniform helicity is approximately one quarter of a period ahead of $\dR$. For the parameters studied here, this mode lies near $50\,\mathrm{GHz}$. Its coupling to a perpendicular field suggests studying the mode by excitation with an oscillating out-of-plane field at comparable frequencies.

\section{Conclusion}

The altermagnetic skyrmion studied here has an internal size coordinate
$\dR=R_A-R_B$ that can be controlled by a uniform out-of-plane magnetic
field. A small $\dR$ contributes linearly to the net magnetic moment and
couples to $B_z$ through the Zeeman term $-\Lambda_R B_z\dR$. Near the
zero-field equilibrium, $\dR$ and $\Nn$ can vary linearly with $B_z$, whereas
the changes in $\Rbar$ and $\Qs$ begin at order $B_z^2$. The field therefore
changes the relative sizes of the two sublattice textures without changing
their mean radius to first order.

The equilibrium size also depends on the relative orientation of the two
sublattice textures. In AM, equal radii do not make the perpendicular
contours coincide. Their difference produces four lobes of local $M_z$
whose integral vanishes at zero field. In the local wall model, this
contour difference gives an excess intersublattice energy that increases
with $\Rbar$ at fixed ellipticity and wall width. The full-energy
calculations give a smaller equilibrium radius in AM than in PAR at the
sampled $\rho>1$.
At $\rho=4$, the PAR radius is $62.23\%$ larger than the
AM radius. Thus, the spatial relation between the two sublattice textures
contributes to the equilibrium size of the composite skyrmion.

The interchanged strong and weak exchange directions also allow a common
directional exchange modulation to produce $\dR\ne0$ at $B_z=0$. The
slope $\chi_\epsilon=(d\dR/d\delta)_0$ vanishes when the two sublattice
exchange tensors are identical and increases over the sampled exchange
contrasts. The perpendicular sublattice geometry therefore allows the
relative radii to be changed by the same exchange modulation on both
sublattices. This mechanism differs from the magnetic-field coupling,
which remains nonzero in the isotropic reference.

Beyond these static deformations, the altermagnetic skyrmion supports an
internal oscillation of $\dR$ coupled to the uniform helicity $\bar\phi$.
A short perpendicular field pulse excites the two coordinates with a phase
difference of approximately one quarter of a period. At $\rho=2$, the LLG
frequency is $49.52\,\mathrm{GHz}$. Using $P$ and $\kappa_{\rm con}$ from
the same fixed-$q$ family and $\kappa_{\bar\phi}$ from a common spin
rotation gives $53.00\,\mathrm{GHz}$.
The predicted frequency and amplitude
ratio are $7.03\%$ and $9.58\%$ above the LLG values, respectively. The
relation $\omega_0=\sqrt{\kappa_{\rm con}\kappa_{\bar\phi}}/|P|$ describes
this motion through the radial and helicity energy coefficients and their
Berry coupling. In the LLG calculations, the AM frequency decreases to
$46.51\,\mathrm{GHz}$ at $\rho=4$. The altermagnetic skyrmion thus supports
a field-excitable internal mode whose frequency varies with the exchange
contrast.

These results suggest a possible use of the internal mode in magnetic
resonant elements operating at microwave frequencies. Calculations under
an oscillating perpendicular field could determine the absorption
spectrum and the amplitude of the oscillating magnetic moment,
providing a basis for evaluating magnetic detection of the mode.
Further work should assess thermal stability and establish whether
lattice strain can tune the oscillation frequency through changes
in the exchange coefficients, anisotropy, and DMI of a specific material.

%

\clearpage
\onecolumngrid

\setcounter{section}{0}
\setcounter{equation}{0}
\setcounter{figure}{0}
\setcounter{table}{0}
\renewcommand{\thesection}{S\arabic{section}}
\renewcommand{\theequation}{S\arabic{equation}}
\renewcommand{\thefigure}{S\arabic{figure}}
\renewcommand{\thetable}{S\arabic{table}}

\begin{center}
{\large\bfseries Supplemental Material for ``Antisymmetric breathing in altermagnetic skyrmions''\par}
\vspace{0.8em}
Yunxi Jiang, Chen Xuan, Xi Chen, Zhikai Wang, Qinfeng Zhu, and Hao Yu
\end{center}

\vspace{0.8em}
\twocolumngrid

\section{Magnetostatic energy and field}
\label{sm:magnetostatics}

We calculate the magnetostatic contribution to the energy and effective
fields in main-text Eqs.~(2) and (3). The magnetization is uniform through
the film thickness $t$ and periodic in the film plane. On the spatial grid,
\begin{equation}
\mathbf M_{ij}=M_s(\mathbf m_{A,ij}+\mathbf m_{B,ij}).
\label{eq:source_S61}
\end{equation}
For a grid field $\mathbf U$, we use the discrete Fourier pair
\[
\begin{aligned}
\widehat{\mathbf U}_{n_xn_y}
 &=\sum_{i=0}^{N_x-1}\sum_{j=0}^{N_y-1}
 \mathbf U_{ij}e^{-2\pi\mathrm i(n_xi/N_x+n_yj/N_y)},\\
\mathbf U_{ij}
 &=\frac{1}{N_xN_y}\sum_{n_x,n_y}\widehat{\mathbf U}_{n_xn_y}
 e^{2\pi\mathrm i(n_xi/N_x+n_yj/N_y)}.
\end{aligned}
\]
Here $N_x,N_y$ are the numbers of grid points, and the integer frequency
indices are interpreted with the usual positive and negative Fourier
wavevectors, $k_x=2\pi n_x/(N_xh)$ and $k_y=2\pi n_y/(N_yh)$.
For $k=|\mathbf k|$, define
\begin{equation}
f_{\rm ms}(kt)=
\begin{cases}
(1-e^{-kt})/(kt),&k>0,\\
1,&k=0.
\end{cases}
\label{eq:source_S62}
\end{equation}
The magnetostatic tensor is
\begin{equation}
\begin{aligned}
N_{\alpha\beta}(\mathbf k)
 &=[1-f_{\rm ms}(kt)]\frac{k_\alpha k_\beta}{k^2},
 &\quad(k>0),\\
N_{zz}(\mathbf k)&=f_{\rm ms}(kt),\\
N_{xz}=N_{zx}&=N_{yz}=N_{zy}=0,\\
N(\mathbf0)&=\operatorname{diag}(0,0,1).
\end{aligned}
\label{eq:demag_kernel_sm}
\end{equation}
Here $\alpha,\beta\in\{x,y\}$.

The field is
\begin{equation}
\mathbf H_{\rm ms}
=-\mu_0\mathcal F^{-1}\!\left[N(\mathbf k)\mathcal F\mathbf M\right],
\label{eq:source_S64}
\end{equation}
and its energy is
\begin{equation}
E_{\rm ms}
=-\frac{h^2t}{2}\sum_{ij}\mathbf M_{ij}\cdot\mathbf H_{{\rm ms},ij}.
\label{eq:demag_energy_sm}
\end{equation}
For either sublattice,
\begin{equation}
-\frac{1}{M_sh^2t}
\frac{\partial E_{\rm ms}}{\partial\mathbf m_{\eta,ij}}
=\mathbf H_{{\rm ms},ij},\qquad \eta=A,B.
\label{eq:demag_field_sm}
\end{equation}
Thus the same $\mathbf H_{\rm ms}$, in tesla, enters both LLG equations.

For $\mathbf m_A(\mathbf r)=-\mathbf m_B(\mathbf r)$ at every point,
\[
\mathbf M\equiv0,\qquad \mathbf H_{\rm ms}\equiv0,\qquad E_{\rm ms}=0.
\]
The Fourier contribution from $M_z$ contains both sublattices:
\[
\begin{aligned}
f_{\rm ms}|\widehat M_z|^2
=M_s^2f_{\rm ms}\bigl(&|\widehat m_{Az}|^2+|\widehat m_{Bz}|^2\\
&+2\operatorname{Re}[\widehat m_{Az}^{\,*}\widehat m_{Bz}]\bigr).
\end{aligned}
\]
The anisotropy coefficient remains $K=K_0/v_c$.
Because $N_{xz}=N_{yz}=0$, the two contributions to the energy are
\[
\begin{aligned}
E_{\rm ms}^{\parallel}
 &=-\frac{h^2t}{2}\sum_{ij}
 \mathbf M_{\parallel,ij}\cdot\mathbf H_{{\rm ms},\parallel,ij},\\
E_{\rm ms}^{z}
 &=-\frac{h^2t}{2}\sum_{ij}M_{z,ij}H_{{\rm ms},z,ij},\\
E_{\rm ms}&=E_{\rm ms}^{\parallel}+E_{\rm ms}^{z}.
\end{aligned}
\]
For AM at $B_z=0$ and $\rho=2$,
\[
E_{\rm ms}^{\parallel}=8.96\times10^{-22}\,\mathrm J,
\qquad E_{\rm ms}^{z}=2.18\times10^{-21}\,\mathrm J.
\]
They contribute approximately $29\%$ and $71\%$ of $E_{\rm ms}$.

\section{Static minimization and contour measurements}
\label{sm:static-methods}

We minimize the total energy $E$ of main-text Eq.~(2).
$E$ is measured
in joules and excludes all numerical constraint terms. The primary grid is
\[
\begin{aligned}
N_x=N_y&=192, & h&=0.5\,\mathrm{nm},\\
N_xh=N_yh&=96\,\mathrm{nm}, & t&=1\,\mathrm{nm}.
\end{aligned}
\]
The boundary conditions are periodic in $x$ and $y$. We use
\[
\begin{aligned}
\bar A&=\sqrt2\times10^{-11}\,\mathrm{J/m},&
 A_0&=2\times10^5\,\mathrm{J/m^3},\\
K&=6\times10^5\,\mathrm{J/m^3},&
 M_s&=5.8\times10^5\,\mathrm{A/m},\\
D&=2.5\times10^{-3}\,\mathrm{J/m^2},&
 A_1&=\bar A\sqrt\rho,\quad A_2=\bar A/\sqrt\rho.
\end{aligned}
\]

The primary static data comprise $\rho=1,1.5,2,2.5,3,3.5,4$ at $B_z=0,\pm0.05\,\mathrm{T}$ for AM and PAR. In AM, the exchange coefficients along $(x,y)$ are $(A_1,A_2)$ on A and $(A_2,A_1)$ on B.
In PAR, both sublattices use $(A_1,A_2)$.

For a grid function $u_{ij}$, define the central differences
\[
D_x^{c}u_{ij}=\frac{u_{i+1,j}-u_{i-1,j}}{2h},\qquad
D_y^{c}u_{ij}=\frac{u_{i,j+1}-u_{i,j-1}}{2h}.
\]
The exchange energy uses nearest-neighbor spin differences, and the DMI
uses $D_x^c,D_y^c$. Its negative energy-density gradient is
\begin{equation}
-\frac{1}{h^2t}\frac{\partial E_D}{\partial\mathbf m_{\eta,ij}}
=2D\begin{pmatrix}
D_x^c m_{\eta z}\\ D_y^c m_{\eta z}\\
-D_x^c m_{\eta x}-D_y^c m_{\eta y}
\end{pmatrix}_{ij}.
\label{eq:source_S60}
\end{equation}
Division by $M_s$ gives the DMI field in tesla.
The forward DMI bonds in main-text Eq.~(1) have the continuum expansions
\[
\begin{aligned}
&-D_0(\mathbf m\times\mathbf m_{i+1,j})\cdot\hat{\mathbf e}_y\\
&\qquad=-D_0a(m_z\partial_xm_x-m_x\partial_xm_z)+O(a^2),\\
&+D_0(\mathbf m\times\mathbf m_{i,j+1})\cdot\hat{\mathbf e}_x\\
&\qquad=-D_0a(m_z\partial_ym_y-m_y\partial_ym_z)+O(a^2).
\end{aligned}
\]
Consequently $D=-D_0a/v_c$. The initial N\'eel orientation is chosen
by evaluating the two opposite radial orientations and selecting the one
with the lower DMI energy. The magnetostatic energy and field of Sec.~\ref{sm:magnetostatics}
are included at every evaluation.

The minimization uses Cartesian energy descent, limited-memory
Broyden--Fletcher--Goldfarb--Shanno (L-BFGS) minimization in angular
coordinates, and a final Cartesian descent. For an unconstrained state,
\[
\begin{aligned}
\mathbf g_{\eta,ij}
 &=-\frac{1}{h^2t}\frac{\partial E}{\partial\mathbf m_{\eta,ij}},\\
\mathbf g_{\eta,ij}^{\perp}
 &=\mathbf g_{\eta,ij}
 -\mathbf m_{\eta,ij}(\mathbf m_{\eta,ij}\cdot\mathbf g_{\eta,ij}),\\
g_{\max}&=\max_{\eta,ij}|\mathbf g_{\eta,ij}^{\perp}|,\\
g_{\rm rms}&=
\left[\frac{1}{2N_xN_y}\sum_{\eta,ij}|\mathbf g_{\eta,ij}^{\perp}|^2\right]^{1/2}.
\end{aligned}
\]
The acceptance conditions are
\[
g_{\max}\le1\,\mathrm{J/m^3},\qquad
g_{\rm rms}\le0.1\,\mathrm{J/m^3},
\]
\[
\max_{\ell\in\mathcal C}R_\eta^{(\ell)}
-\min_{\ell\in\mathcal C}R_\eta^{(\ell)}<10^{-4}\,\mathrm{nm},
\qquad \eta=A,B.
\]
The set $\mathcal C$ contains the last five recorded Cartesian states,
recorded every 200 steps, and the final state.

The spatial origin is fixed at the center of the grid:
\[
x_i=h\left(i-\frac{N_x-1}{2}\right),\qquad
y_j=h\left(j-\frac{N_y-1}{2}\right).
\]
Along a ray at angle $\varphi$, the bilinearly interpolated values of
$m_{\eta z}$ are sampled at $r_\ell=0.10h\,\ell$. Let $z_\ell$ and
$z_{\ell+1}$ bracket the first sign change along this ray. The contour
crossing is obtained from
\[
r_\eta(\varphi)
=r_\ell-\frac{z_\ell}{z_{\ell+1}-z_\ell}(0.10h).
\]
The measured axial distances are
\[
\begin{aligned}
b_\eta&=\frac{r_\eta(0)+r_\eta(\pi)}{2},\\
a_\eta&=\frac{r_\eta(\pi/2)+r_\eta(3\pi/2)}{2},\\
R_\eta&=\sqrt{a_\eta b_\eta},\qquad
\varepsilon_\eta=\frac{b_\eta-a_\eta}{b_\eta+a_\eta}.
\end{aligned}
\]
These give the main-text coordinates
\[
\begin{aligned}
\Rbar&=\frac{R_A+R_B}{2},& \dR&=R_A-R_B,\\
\Qs&=\frac{\varepsilon_A-\varepsilon_B}{2},&
\Nn&=\frac{\varepsilon_A+\varepsilon_B}{2}.
\end{aligned}
\]
The complete contour is sampled at
$\varphi_j=2\pi j/180$, $j=0,\ldots,180$.
The values at $j=0$ and $j=180$ refer to the same direction.
Calculations requiring distinct directions use $j=0,\ldots,179$.

\section{Symmetry under $\mathcal P$}
\label{sm:symmetry}

Magnetization transforms as an axial vector. Reflection about $x=y$ combined with sublattice exchange therefore acts as Eq.~(\ref{eq:Pdef}).

\begin{equation}
\begin{aligned}
\mathcal{P}:\quad
\mathbf{m}'_A(x,y)&=-\mathcal{R}\,\mathbf{m}_B(y,x),\\
\mathbf{m}'_B(x,y)&=-\mathcal{R}\,\mathbf{m}_A(y,x).
\end{aligned}
\label{eq:Pdef}
\end{equation}
Here $\mathcal R(m_x,m_y,m_z)=(m_y,m_x,m_z)$. Primed fields are functions of $(x,y)$, while the unprimed fields on the right are evaluated at $(X,Y)=(y,x)$. The derivatives therefore transform by the chain rule as follows.

\begin{equation}
\begin{aligned}
\partial_x\big[F(y,x)\big]&=\big[\partial_YF\big](y,x),\\
\partial_y\big[F(y,x)\big]&=\big[\partial_XF\big](y,x).
\end{aligned}
\label{eq:chain}
\end{equation}
that is, $\partial_x\leftrightarrow\partial_Y$ and $\partial_y\leftrightarrow\partial_X$.

\noindent Exchange. Using Eqs.~(\ref{eq:Pdef}) and (\ref{eq:chain}), and
$|\mathcal{R}\mathbf{v}|=|\mathbf{v}|$,
\begin{equation}
\begin{aligned}
A_1|\partial_x\mathbf{m}'_A|^2&=A_1|\partial_Y\mathbf{m}_B|^2,\\
A_2|\partial_y\mathbf{m}'_A|^2&=A_2|\partial_X\mathbf{m}_B|^2,\\
A_2|\partial_x\mathbf{m}'_B|^2&=A_2|\partial_Y\mathbf{m}_A|^2,\\
A_1|\partial_y\mathbf{m}'_B|^2&=A_1|\partial_X\mathbf{m}_A|^2.
\end{aligned}
\label{eq:source_S13}
\end{equation}
The four exchange terms reproduce the original sum because the coefficients $A_1$ and $A_2$ are interchanged between the sublattices.

\noindent Intersublattice coupling.
$A_0\mathbf{m}'_A\!\cdot\!\mathbf{m}'_B
=A_0(-\mathcal{R}\mathbf{m}_B)\!\cdot\!(-\mathcal{R}\mathbf{m}_A)
=A_0\mathbf{m}_B\!\cdot\!\mathbf{m}_A$, invariant, since $\mathcal{R}$ is orthogonal and
the two sign changes cancel.

\noindent Anisotropy. $m'_{Az}=-m_{Bz}$ and $m'_{Bz}=-m_{Az}$. The squared components are interchanged, $m_{Az}'{}^2=m_{Bz}^2$ and $m_{Bz}'{}^2=m_{Az}^2$, so their sum is unchanged.

\noindent Dzyaloshinskii-Moriya interaction. Write
$w_D=D\big[m_z(\partial_xm_x+\partial_ym_y)-(m_x\partial_xm_z+m_y\partial_ym_z)\big]$.
Using Eqs.~(\ref{eq:Pdef}) and (\ref{eq:chain}), and noting that $\mathcal{R}$ exchanges
the in-plane components while the overall sign is squared in every term,
\begin{align}
\partial_xm'_{Ax}
&=\partial_x\big[-m_{By}(y,x)\big]\notag\\
&=-\big[\partial_Ym_{By}\big](y,x),\label{eq:dmi_dx_sm}\\
\partial_ym'_{Ay}
&=\partial_y\big[-m_{Bx}(y,x)\big]\notag\\
&=-\big[\partial_Xm_{Bx}\big](y,x).\label{eq:dmi_dy_sm}
\end{align}
so
\begin{equation}
\begin{aligned}
m'_{Az}\big(\partial_xm'_{Ax}+\partial_ym'_{Ay}\big)
&=\big[m_{Bz}\partial_Ym_{By}\\
&\quad+m_{Bz}\partial_Xm_{Bx}\big](y,x)\\
&=\big[m_{Bz}\nabla\!\cdot\!\mathbf{m}_{B\parallel}\big](y,x).
\end{aligned}
\label{eq:source_S16}
\end{equation}
and likewise
\begin{equation}
\begin{aligned}
m'_{Ax}\partial_xm'_{Az}+m'_{Ay}\partial_ym'_{Az}
&=\big[(\mathbf{m}_{B\parallel}\!\cdot\!\nabla)m_{Bz}\big](y,x).
\end{aligned}
\label{eq:source_S17}
\end{equation}

Equations~(\ref{eq:dmi_dx_sm})--(\ref{eq:source_S17}) give
\[
\begin{aligned}
w_D[\mathbf m'_A](x,y)&=w_D[\mathbf m_B](y,x),\\
w_D[\mathbf m'_B](x,y)&=w_D[\mathbf m_A](y,x).
\end{aligned}
\]
Since $dx\,dy=dX\,dY$ for $(X,Y)=(y,x)$,
\[
E_D[\mathbf m'_A]+E_D[\mathbf m'_B]
=E_D[\mathbf m_A]+E_D[\mathbf m_B].
\]

\noindent Zeeman energy. The transformed magnetization satisfies
\[
M'_z(x,y)=-M_z(y,x),\qquad \mu'_z=-\mu_z.
\]
Therefore
\[
\begin{aligned}
E_Z[\mathcal P\mathbf m;B_z]
 &=-B_z(-\mu_z)=-E_Z[\mathbf m;B_z],\\
E_Z[\mathcal P\mathbf m;-B_z]
 &=-(-B_z)(-\mu_z)=E_Z[\mathbf m;B_z].
\end{aligned}
\]
Here $\mathbf m$ denotes the pair $(\mathbf m_A,\mathbf m_B)$.

\noindent Magnetostatic energy. With $\mathcal R(x,y,z)=(y,x,z)$,
\[
\widehat{\mathbf M}'(\mathbf k)
=-\mathcal R\widehat{\mathbf M}(\mathcal R\mathbf k),\qquad
N(\mathcal R\mathbf k)=\mathcal R N(\mathbf k)\mathcal R^T.
\]
Consequently,
\[
\widehat{\mathbf M}'(\mathbf k)^\dagger N(\mathbf k)
\widehat{\mathbf M}'(\mathbf k)
=
\widehat{\mathbf M}(\mathcal R\mathbf k)^\dagger
N(\mathcal R\mathbf k)\widehat{\mathbf M}(\mathcal R\mathbf k).
\]
The sum over wavevectors in the square periodic cell gives
$E_{\rm ms}[\mathcal P\mathbf m]=E_{\rm ms}[\mathbf m]$.
Here $\dagger$ denotes complex conjugate transpose.

\noindent Collective coordinates. The contour transforms as
\begin{equation}
\begin{aligned}
r'_A(\varphi)&=r_B(\pi/2-\varphi),&
r'_B(\varphi)&=r_A(\pi/2-\varphi),\\
a'_A&=b_B,& b'_A&=a_B,\\
a'_B&=b_A,& b'_B&=a_A.
\end{aligned}
\label{eq:source_S18}
\end{equation}
Thus $R'_A=R_B$, $R'_B=R_A$,
$\varepsilon'_A=-\varepsilon_B$, and
$\varepsilon'_B=-\varepsilon_A$, giving
\begin{equation}
(\Rbar,\Qs,\dR,\Nn)
\longmapsto(\Rbar,\Qs,-\dR,-\Nn).
\label{eq:source_S19}
\end{equation}
Combining this result with the field reversal gives
\[
E(\Rbar,\Qs,\dR,\Nn;B_z)
=E(\Rbar,\Qs,-\dR,-\Nn;-B_z).
\]
For a term $\dR^{n_R}\Nn^{n_e}B_z^{n_B}$ in the expansion at fixed $\Rbar,\Qs$, with nonnegative integer exponents,
\[
n_R+n_e+n_B\ \text{must be even}.
\]
In particular, $B_z\dR$, $B_z\Nn$, and $\dR\Nn$ are allowed.
For the smooth equilibrium branch connected to the zero-field state
$\dR=\Nn=0$,
\[
\begin{aligned}
\dR(-B_z)&=-\dR(B_z),&\Nn(-B_z)&=-\Nn(B_z),\\
\Rbar(-B_z)&=\Rbar(B_z),&\Qs(-B_z)&=\Qs(B_z).
\end{aligned}
\]
Section~\ref{sm:ellipticity-coupling} estimates the coefficient $\Lambda_{\Nn}$ of the field coupling
to $\Nn$.

A rotation by $\pi/2$ about $z$, combined with sublattice exchange, acts as
\[
\begin{aligned}
\mathbf m'_A(\mathbf r)
 &=R_z(\pi/2)\mathbf m_B\!\left(R_z(-\pi/2)\mathbf r\right),\\
\mathbf m'_B(\mathbf r)
 &=R_z(\pi/2)\mathbf m_A\!\left(R_z(-\pi/2)\mathbf r\right).
\end{aligned}
\]
Here $R_z$ rotates both position and spin vectors about $z$.
This transformation gives
\[
\begin{aligned}
(m_{Az}^{\rm core},m_{Bz}^{\rm core})&:(+1,-1)\mapsto(-1,+1),\\
(\dR,\Nn,\mu_z,B_z)&\mapsto(-\dR,-\Nn,\mu_z,B_z).
\end{aligned}
\]
The energy is unchanged at the same $B_z$, but the core polarities are
reversed. The field-parity relations above apply to the fixed core
polarities $(+1,-1)$ used in the calculations.

For the quadratic energy in main-text Eq.~(8), stationarity at fixed
$\Rbar$ and $\Qs$ gives
\[
\begin{pmatrix}\kappa_0&\zeta\\\zeta&\kappa_{\Nn}\end{pmatrix}
\begin{pmatrix}\dR\\\Nn\end{pmatrix}
=B_z\begin{pmatrix}\Lambda_R\\\Lambda_{\Nn}\end{pmatrix}.
\]
For $\kappa_{\Nn}>0$, eliminating $\Nn$ gives
\[
\begin{aligned}
\kappa&=\kappa_0-\frac{\zeta^2}{\kappa_{\Nn}},\\
\Nn&=\frac{\Lambda_{\Nn}B_z-\zeta\dR}{\kappa_{\Nn}},\\
\dR&=\frac{\Lambda_R-\zeta\Lambda_{\Nn}/\kappa_{\Nn}}{\kappa}B_z
\equiv\chi B_z.
\end{aligned}
\]
At $B_z=0$, with $\dR$ prescribed,
\[
\Nn=-\frac{\zeta}{\kappa_{\Nn}}\dR,
\qquad
\Delta E(\dR)=\frac12\kappa\dR^2+O(\dR^4),
\]
where $\Delta E$ is the energy relative to $\dR=0$ after minimizing
$\Nn$ at the same $\Rbar,\Qs$. The quadratic form is positive when
\[
\kappa_{\Nn}>0,\qquad
\kappa_0\kappa_{\Nn}-\zeta^2>0,
\]
which is equivalent to $\kappa_{\Nn}>0$ and $\kappa>0$.

\section{Small-field slopes and numerical checks}
\label{sm:field-slopes}

The zero-field slope in main-text Eq.~(10) is estimated from independently
minimized states at $B_z=\pm B_1$. We calculate
\[
\begin{aligned}
\mu_z(B_z)&=M_sth^2\sum_{ij}
[m_{Az,ij}(B_z)+m_{Bz,ij}(B_z)],\\
\Delta X&=X(B_1)-X(-B_1).
\end{aligned}
\]
For $B_1=0.05\,\mathrm T$,
\begin{equation}
\chi=\left.\frac{d\dR}{dB_z}\right|_0
\simeq\frac{\Delta\dR}{2B_1},
\label{eq:response_sm}
\end{equation}
\[
\Lambda_{\rm path}=\frac{\Delta\mu_z}{\Delta\dR}.
\]
For an analytic odd dependence
$\dR(B_z)=\chi B_z+c_3B_z^3+O(B_z^5)$,
\[
\frac{\Delta\dR}{2B_1}=\chi+c_3B_1^2+O(B_1^4).
\]
Thus the difference at a specified $B_1$ estimates the derivative at zero.
All spin variables can change between the two states defining
$\Lambda_{\rm path}$.
For the $h=0.5\,\mathrm{nm}$ grid,
\[
\begin{array}{c|c|c|c}
\rho&\text{geometry}&\chi\ (\mathrm{nm/T})&
\Lambda_{\rm path}\ (\mathrm{J\,T^{-1}m^{-1}})\\ \hline
2&\mathrm{AM}&3.08&3.73\times10^{-11}\\
4&\mathrm{AM}&3.49&3.50\times10^{-11}\\
4&\mathrm{PAR}&3.70&4.89\times10^{-11}
\end{array}
\]
The same differences are evaluated for the seven $\rho$ values and both
geometries specified in Sec.~\ref{sm:static-methods}. The zero-field states give $\Rbar$ in
main-text Table~I.
The states at $\pm0.05\,\mathrm T$ give the listed
estimates of $\chi$.

At $\rho=2$, define the even difference
\[
\Delta e_s^{\rm even}(B_1)
=\frac{\Qs(B_1)+\Qs(-B_1)}{2}-\Qs(0).
\]
The additional small-field calculations give
\[
\begin{array}{c|r|r}
B_1\ (\mathrm T)&\Delta e_s^{\rm even}&
\Delta e_s^{\rm even}/B_1^2\ (\mathrm{T^{-2}})\\ \hline
0.0125&2.79\times10^{-6}&1.79\times10^{-2}\\
0.025&-7.23\times10^{-5}&-1.16\times10^{-1}
\end{array}
\]

For $\rho=2$ in the same $96\times96\,\mathrm{nm}^2$ periodic cell,
using $B_z=\pm0.05\,\mathrm T$,
\[
\begin{array}{c|c}
h\ (\mathrm{nm})&\chi_{\parallel}-\chi_{\rm AM}\ (\mathrm{nm/T})\\ \hline
0.50&1.83\times10^{-2}\\
0.25&6.17\times10^{-2}
\end{array}
\]
PAR has the larger estimate of $\chi$ on both grids, while the magnitude of the difference
remains sensitive to $h$.

\section{Contour expansion and local magnetization}
\label{sm:contours}

For the elliptical contour of sublattice $\eta$,
\[
\frac{x^2}{b_\eta^2}+\frac{y^2}{a_\eta^2}=1,\qquad
x=r\cos\varphi,\quad y=r\sin\varphi.
\]
Its polar radius is
\begin{equation}
r_\eta(\varphi)
=\frac{a_\eta b_\eta}
{\sqrt{a_\eta^2\cos^2\varphi+b_\eta^2\sin^2\varphi}}.
\label{eq:source_S1}
\end{equation}
Using the main-text definitions
$R_\eta=\sqrt{a_\eta b_\eta}$ and
$\varepsilon_\eta=(b_\eta-a_\eta)/(b_\eta+a_\eta)$ gives
\begin{equation}
a_\eta=R_\eta\sqrt{\frac{1-\varepsilon_\eta}{1+\varepsilon_\eta}},
\qquad
b_\eta=R_\eta\sqrt{\frac{1+\varepsilon_\eta}{1-\varepsilon_\eta}},
\label{eq:source_S2}
\end{equation}
\begin{equation}
\frac{r_\eta(\varphi)}{R_\eta}
=\frac{\sqrt{1-\varepsilon_\eta^2}}
{\sqrt{1-2\varepsilon_\eta\cos2\varphi+\varepsilon_\eta^2}}.
\label{eq:source_S3}
\end{equation}
Expanding in $\varepsilon_\eta$,
\begin{equation}
\begin{aligned}
\frac{r_\eta}{R_\eta}
 &=1+\varepsilon_\eta\cos2\varphi-\varepsilon_\eta^2
 +\frac32\varepsilon_\eta^2\cos^22\varphi+O(\varepsilon_\eta^3)\\
 &=1+\varepsilon_\eta\cos2\varphi
 -\frac{\varepsilon_\eta^2}{4}
 +\frac{3\varepsilon_\eta^2}{4}\cos4\varphi
 +O(\varepsilon_\eta^3).
\end{aligned}
\label{eq:source_S4}
\end{equation}
The exact ellipse obeys
\[
\frac12\int_0^{2\pi}r_\eta^2\,d\varphi=\pi a_\eta b_\eta=\pi R_\eta^2.
\]
In the expansion, the mean contribution from the terms of order
$\varepsilon_\eta^2$ in $r_\eta^2/R_\eta^2$ is
\[
\left\langle -\frac{\varepsilon_\eta^2}{2}
+\varepsilon_\eta^2\cos^22\varphi
+\frac32\varepsilon_\eta^2\cos4\varphi\right\rangle_\varphi=0,
\]
where $\langle X\rangle_\varphi=(2\pi)^{-1}\int_0^{2\pi}X\,d\varphi$.
For the numerical contours, $R_\eta$ is measured from the axial
crossings as specified in Sec.~\ref{sm:static-methods}.

For a centered contour symmetric about the $x$ and $y$ axes,
\[
\frac{r_\eta(\varphi)}{R_\eta}
=c_{0,\eta}+\sum_{n\ge1}c_{2n,\eta}\cos(2n\varphi),
\]
\[
\begin{aligned}
c_{0,\eta}&=\frac{1}{2\pi}\int_0^{2\pi}
\frac{r_\eta(\varphi)}{R_\eta}\,d\varphi,\\
c_{2n,\eta}&=\frac1\pi\int_0^{2\pi}
\frac{r_\eta(\varphi)}{R_\eta}\cos(2n\varphi)\,d\varphi.
\end{aligned}
\]
For an ellipse,
\[
\left|\frac{c_{4,\eta}}{c_{2,\eta}}\right|
=\frac34|\varepsilon_\eta|+O(\varepsilon_\eta^3),\qquad
\left|\frac{c_{6,\eta}}{c_{2,\eta}}\right|
=\frac58\varepsilon_\eta^2+O(\varepsilon_\eta^4).
\]
For the zero-field AM state invariant under $\mathcal P$,
\[
R_A=R_B=\Rbar,\qquad
c_{2n,B}=(-1)^n c_{2n,A},
\]
\[
\frac{u(\varphi)}{\Rbar}
=2\sum_{j\ge0}c_{4j+2,A}\cos[(4j+2)\varphi].
\]
Thus the common $\cos4\varphi$ term cancels, whereas
$\cos2\varphi$ and $\cos6\varphi$ are allowed in $u$.

For the approximation used in main-text Eq.~(11),
\begin{equation}
\begin{aligned}
m_{Az}(r,\varphi)&=f\!\left(\frac r{r_A(\varphi)}\right),\\
m_{Bz}(r,\varphi)&=-f\!\left(\frac r{r_B(\varphi)}\right),
\end{aligned}
\label{eq:source_S20}
\end{equation}
where $f(0)=1$, $f(1)=0$, and $f(\infty)=-1$.
At fixed $r$,
\[
\left.\frac{\partial f(r/\ell)}{\partial\ell}\right|_{\ell=\Rbar}
=-\frac r{\Rbar^2}f'\!\left(\frac r{\Rbar}\right).
\]
To first order in $r_A(\varphi)-\Rbar$ and
$r_B(\varphi)-\Rbar$,
\begin{equation}
\begin{aligned}
M_z(r,\varphi)
 &=M_s\left[f\!\left(\frac r{r_A}\right)
             -f\!\left(\frac r{r_B}\right)\right]\\
 &\simeq M_sp(r)u(\varphi),\\
p(r)&=-\frac r{\Rbar^2}f'\!\left(\frac r{\Rbar}\right).
\end{aligned}
\label{eq:Mz_sm}
\end{equation}
The function $p$ has units of inverse length.
It is positive where $f$
decreases through the wall.
Separately, the first-order expansion of each elliptical contour gives
\begin{equation}
\begin{aligned}
u(\varphi)&=r_A(\varphi)-r_B(\varphi)\\
&\simeq\dR+(R_A\varepsilon_A-R_B\varepsilon_B)\cos2\varphi\\
&=\dR+(2\Rbar\Qs+\dR\Nn)\cos2\varphi.
\end{aligned}
\label{eq:u_sm}
\end{equation}

At zero field, $\dR=\Nn=0$, so
\[
u(\varphi)\simeq 2\Rbar\Qs\cos2\varphi.
\]
For the calculated $\rho=2$ state,
\[
\begin{aligned}
u(\varphi)&\simeq(1.29\,\mathrm{nm})\cos2\varphi,\\
\frac{2\Rbar\Qs}{\lambda}&=2.68\times10^{-1},
\qquad \lambda=\sqrt{\bar A/K},\\
\frac{\max_{\mathbf r}|M_z(\mathbf r)|}{M_s}&=4.65\times10^{-1},\\
\frac{\max_{\mathbf r}|M_z(\mathbf r)|/M_s}
     {2\Rbar\Qs/\lambda}&=1.73.
\end{aligned}
\]
The numerical moments in Sec.~\ref{sm:moments} are evaluated from the complete spin
configurations, rather than from the first-order expression for $M_z$.

\section{Net moment and higher magnetic moments}
\label{sm:moments}

The moments used below are
\begin{equation}
\mu_z=t\int d^2r\,M_z,\qquad
\mathcal M_z^{(3)}=t\int d^2r\,(x^2-y^2)M_z.
\label{eq:source_S23}
\end{equation}
Here $x^2-y^2=r^2\cos2\varphi$,
$[\mu_z]=\mathrm{A\,m^2}=\mathrm{J/T}$, and
$[\mathcal M_z^{(3)}]=\mathrm{A\,m^4}$.
The superscript 3 counts one magnetization component and two position
factors in this Cartesian moment \cite{SM-PhysRevLett.133.196701,SM-xtcd-t47t}.

Assume the radial moments converge and define
\[
W_n=\int_0^\infty r^{1+n}p(r)\,dr,
\qquad [W_n]=\mathrm{m}^{n+1}.
\]
Using $\int_0^{2\pi}\cos2\varphi\,d\varphi=0$ in the first-order
magnetization gives
\begin{equation}
\mu_z=2\pi M_st W_0\dR\equiv\Lambda_R\dR.
\label{eq:mu_sm}
\end{equation}
At $\dR=\Nn=0$, retaining the leading term in $\Qs$ and using
$\int_0^{2\pi}\cos^22\varphi\,d\varphi=\pi$ gives
\begin{equation}
\mathcal M_z^{(3)}\simeq2\pi M_st W_2\Rbar\Qs.
\label{eq:quad_sm}
\end{equation}

For an area estimate, replace the smooth radial function by
\[
f(\xi)=
\begin{cases}
+1,&0\le\xi<1,\\
-1,&\xi>1.
\end{cases}
\]
$f'(\xi)=-2\delta_{\rm D}(\xi-1)$, where $\delta_{\rm D}$ is the
Dirac delta distribution. Hence
\[
p(r)=\frac{2r}{\Rbar}\delta_{\rm D}(r-\Rbar),\qquad
W_0=2\Rbar,\quad W_2=2\Rbar^3.
\]
For exact elliptical interiors $\mathcal D_A,\mathcal D_B$,
\[
\begin{aligned}
\mu_z
 &=2M_st\left(\int_{\mathcal D_A}d^2r
              -\int_{\mathcal D_B}d^2r\right)\\
 &=2\pi M_st(R_A^2-R_B^2)=4\pi M_st\Rbar\dR.
\end{aligned}
\]
This gives $\Lambda_R=4\pi M_st\Rbar$ in main-text Eq.~(16).
For the same piecewise profile,
\[
\begin{aligned}
\mathcal M_z^{(3)}
 &=\frac{\pi M_st}{2}\left[
 R_A^2(b_A^2-a_A^2)-R_B^2(b_B^2-a_B^2)\right]\\
 &=2\pi M_st\left[
 \frac{R_A^4\varepsilon_A}{1-\varepsilon_A^2}
 -\frac{R_B^4\varepsilon_B}{1-\varepsilon_B^2}\right].
\end{aligned}
\]
We express deviations from the leading area estimates by
\begin{equation}
\Lambda_R=4\pi c_\mu M_st\Rbar,\qquad
\mathcal M_z^{(3)}=4\pi c_Q M_st\Rbar^4\Qs.
\label{eq:Lambda_sm}
\end{equation}
The second definition is used for zero-field states with $\Qs\ne0$.
For the piecewise profile, $c_\mu=1$. At $\dR=\Nn=0$,
\begin{equation}
\mathcal M_z^{(3)}
=4\pi M_st\Rbar^4\frac{\Qs}{1-\Qs^2}
=4\pi M_st\Rbar^4\Qs+O(\Qs^3),
\label{eq:source_S27}
\end{equation}
so $c_Q=(1-\Qs^2)^{-1}$ for exact ellipses, and $c_Q=1$ to leading
order in $\Qs$.
\[
\left.
\frac{\partial^2\mathcal M_z^{(3)}}{\partial\dR\,\partial\Nn}
\right|_{\dR=\Nn=0}
=8\pi M_st\Rbar^3\frac{1+\Qs^2}{(1-\Qs^2)^2},
\]
with $\Rbar,\Qs$ fixed. Its leading coefficient at small $\Qs$ is
$8\pi M_st\Rbar^3$, multiplying $\dR\Nn$.

For each computed zero-field state, we evaluate
\[
c_Q=\frac{\mathcal M_z^{(3)}}{4\pi M_st\Rbar^4\Qs}.
\]
The values are
\[
\begin{array}{c|rrrrrr}
\rho&1.5&2&2.5&3&3.5&4\\ \hline
c_Q&5.53&5.94&6.39&6.80&7.15&7.46
\end{array}
\]
The radial weight in $\mathcal M_z^{(3)}$ contains two more powers of
$r$ than that in $\mu_z$.
Magnetization outside the central part of the
wall therefore contributes more strongly to $c_Q$. Both $c_Q$ and
$\Rbar$ change along this scan. At $\rho=1$,
$\mathcal M_z^{(3)}=0$.
For AM at $\rho=2$,
\[
\begin{aligned}
\frac{\Lambda_{\rm path}}{4\pi M_st\Rbar}&=1.32,\\
\Lambda_{\rm path}&=\frac{\mu_z(B_1)-\mu_z(-B_1)}
{\dR(B_1)-\dR(-B_1)},\qquad B_1=0.05\,\mathrm T.
\end{aligned}
\]
This ratio uses the field-minimized states of Sec.~\ref{sm:field-slopes}.

For an exact ellipse and any common scaled function $f$, use
$x=b_\eta\xi\cos\vartheta$ and
$y=a_\eta\xi\sin\vartheta$. Then
\[
\frac r{r_\eta(\varphi)}=\xi,\qquad
d^2r=a_\eta b_\eta\,\xi\,d\xi\,d\vartheta,
\]
\[
\begin{aligned}
\int d^2r\,[1+s_\eta m_{\eta z}]
&=2\pi R_\eta^2\int_0^\infty\xi[1+f(\xi)]\,d\xi,\\
s_A&=+1,\qquad s_B=-1.
\end{aligned}
\]
Consequently,
\[
\begin{aligned}
\mu_z&=2\pi M_st(R_A^2-R_B^2)
       \int_0^\infty\xi[1+f(\xi)]\,d\xi,\\
\left.\frac{\partial\mu_z}{\partial\Nn}\right|_{R_A,R_B,f}&=0.
\end{aligned}
\]
Thus $\Lambda_{\Nn}=0$ within this specified family, as used in
main-text Eq.~(17).
The numerical moment instead uses the complete spin configuration:
\begin{equation}
\mathcal M_z^{(3)}
=M_sth^2\sum_{ij}(m_{Az,ij}+m_{Bz,ij})(x_i^2-y_j^2).
\label{eq:multipole_sum}
\end{equation}

For a radial trial state, write
\[
\mathbf m_{A\parallel}=f_\parallel(r/r_A)\hat{\mathbf r},\qquad
\mathbf m_{B\parallel}=-f_\parallel(r/r_B)\hat{\mathbf r},
\]
where $f_\parallel$ is the signed radial component of the reference unit
spin and $f_\parallel^2+f^2=1$. To first order in the contour displacements,
\[
\begin{aligned}
\mathbf M_\parallel&\simeq\widetilde p(r)u(\varphi)\hat{\mathbf r},\\
\widetilde p(r)&=-\frac{M_sr}{\Rbar^2}
 f_\parallel'\!\left(\frac r{\Rbar}\right),
\qquad [\widetilde p]=\mathrm{A/m^2}.
\end{aligned}
\]
The position-weighted in-plane moment is
\begin{equation}
\begin{aligned}
\mathcal M_{ij}^{(2)}&=t\int d^2r\,r_iM_j,\qquad i,j\in\{x,y\},\\
[\mathcal M_{ij}^{(2)}]&=\mathrm{A\,m^3}.
\end{aligned}
\label{eq:source_S28}
\end{equation}
At zero field, $u\simeq2\Rbar\Qs\cos2\varphi$. The angular integrals give
\[
t\int d^2r\,\mathbf M_\parallel=0,
\]
\[
\begin{pmatrix}
\mathcal M_{xx}^{(2)}&\mathcal M_{xy}^{(2)}\\
\mathcal M_{yx}^{(2)}&\mathcal M_{yy}^{(2)}
\end{pmatrix}
\simeq
\pi t\Rbar\Qs\!\int_0^\infty r^2\widetilde p(r)\,dr
\begin{pmatrix}1&0\\0&-1\end{pmatrix}.
\]
Direct integration of the computed spin configurations gives
\[
\begin{array}{c|r}
\rho&-\mathcal M_{xx}^{(2)}\ (10^{-28}\,\mathrm{A\,m^3})\\ \hline
1.5&1.09\\2&1.82\\2.5&2.34\\3&2.76\\3.5&3.10\\4&3.41
\end{array}
\]
with $\mathcal M_{yy}^{(2)}=-\mathcal M_{xx}^{(2)}$ to the reported
precision. The ratio to the contour-displacement scale is
\begin{equation}
\frac{-\mathcal M_{xx}^{(2)}}{2\Rbar\Qs/\lambda}
=\begin{cases}
6.81\times10^{-28}\,\mathrm{A\,m^3},&\rho=1.5,\\
6.87\times10^{-28}\,\mathrm{A\,m^3},&\rho=4.
\end{cases}
\label{eq:source_S29}
\end{equation}
The endpoint values differ by approximately $1.0\%$.
At $\rho=1$, $\mathbf M\equiv0$ and $\mathcal M_{ij}^{(2)}=0$.
For a field-dependent state, the sum of the diagonal components is
\[
\mathcal M_{xx}^{(2)}+\mathcal M_{yy}^{(2)}
=t\int d^2r\,(xM_x+yM_y),
\]
\[
[\mathcal M_{xx}^{(2)}+\mathcal M_{yy}^{(2)}](-B_z)
=-[\mathcal M_{xx}^{(2)}+\mathcal M_{yy}^{(2)}](B_z).
\]
For an isolated skyrmion and observation distance $d$ much larger than
its spatial extent, the rank-2 and rank-3 field contributions have radial
dependences $d^{-4}$ and $d^{-5}$, respectively.

\section{Field coupling to the uniform ellipticity}
\label{sm:ellipticity-coupling}

For AM at $B_z=0$ and $\rho=2$, we vary the integral-based ellipticity
constraint and adjust the area constraint so that the measured radius
difference remains near zero. The constrained quantities are
\begin{equation}
\begin{aligned}
q&=\sqrt{S_A/\pi}-\sqrt{S_B/\pi},\\
\widetilde e_u&=\frac{\widetilde\varepsilon_A+
                              \widetilde\varepsilon_B}{2},\qquad
\widetilde\varepsilon_\eta=\frac{\pi T_\eta}{S_\eta^2}.
\end{aligned}
\label{eq:source_S32}
\end{equation}
Here
\[
\begin{aligned}
S_\eta&=\frac{h^2}{2}\sum_{ij}(1+s_\eta m_{\eta z,ij}),\\
T_\eta&=\frac{h^2}{2}\sum_{ij}(x_i^2-y_j^2)(1+s_\eta m_{\eta z,ij}),\\
s_A&=+1,\qquad s_B=-1.
\end{aligned}
\]

The imposed targets and acceptance conditions are
\[
\widetilde e_u^{\,*}\in\{0,\pm0.01,\pm0.02\},
\]
\[
\begin{aligned}
|q-q^*|&\le0.002\,\mathrm{nm},\\
|\widetilde e_u-\widetilde e_u^{\,*}|&\le2\times10^{-5},\\
|\dR|&\le0.002\,\mathrm{nm}.
\end{aligned}
\]

The five measured values used in the fits are
\[
\Nn\simeq0,\quad\pm8.92\times10^{-4},\quad\pm1.77\times10^{-3}.
\]
We fit $\mu_z$ and the physical energy against these measured values
of $\Nn$. The resulting coefficients
estimate the fixed-coordinate derivatives in main-text Eq.~(8).

The coefficients in main-text Eq.~(8) are defined at
$B_z=\dR=\Nn=0$, with $\Rbar,\Qs$ fixed:
\[
\Lambda_{\Nn}
=\left.\frac{\partial\mu_z}{\partial\Nn}\right|_0,\qquad
\kappa_{\Nn}
=\left.\frac{\partial^2E}{\partial\Nn^2}\right|_0.
\]
Let $e_j$ denote the measured $\Nn$ in the five calibrated states.
The unweighted fits are
\[
(\mu_{\rm off},\ell_\mu)
=\underset{c,\ell}{\operatorname{argmin}}
\sum_{j=1}^{5}(\mu_{z,j}-c-\ell e_j)^2,
\]
\[
(E_{\rm off},b_E,k_e)
=\underset{c,b,k}{\operatorname{argmin}}
\sum_{j=1}^{5}\left(E_j-c-be_j-\frac{k}{2}e_j^2\right)^2.
\]
The fit coefficients estimate the main-text derivatives as
$\Lambda_{\Nn}\simeq\ell_\mu$ and $\kappa_{\Nn}\simeq k_e$.
The calibrated-state fits give the estimates
\begin{equation}
\Lambda_{\Nn}\simeq1.00809327\times10^{-19}\,\mathrm{J/T},
\label{eq:source_S33}
\end{equation}
\begin{equation}
\kappa_{\Nn}\simeq1.59355329\times10^{-17}\,\mathrm J.
\label{eq:source_S34}
\end{equation}

Minimizing the quadratic energy in main-text Eq.~(8) at prescribed
$\dR$ gives
\begin{equation}
0=\kappa_{\Nn}\Nn+\zeta\dR-\Lambda_{\Nn}B_z,
\qquad
\Nn=\frac{\Lambda_{\Nn}}{\kappa_{\Nn}}B_z
     -\frac{\zeta}{\kappa_{\Nn}}\dR.
\label{eq:source_S35}
\end{equation}
At unconstrained small-field equilibrium,
$\dR=\chi B_z+O(B_z^3)$, and therefore
\[
\Nn=\frac{\Lambda_{\Nn}-\zeta\chi}{\kappa_{\Nn}}B_z+O(B_z^3).
\]
For the field pair $B_z=\pm0.05\,\mathrm T$,
\[
\begin{aligned}
\frac{\Delta\Nn}{\Delta B_z}&=-4.42889\times10^{-3}\,\mathrm{T^{-1}},\\
\frac{\Lambda_{\Nn}}{\kappa_{\Nn}}&=6.32607\times10^{-3}\,\mathrm{T^{-1}}.
\end{aligned}
\]
Using the quadratic model to compare these values gives
\[
-\frac{\zeta}{\kappa_{\Nn}}
\frac{\Delta\dR}{\Delta B_z}
\simeq
\frac{\Delta\Nn}{\Delta B_z}
-\frac{\Lambda_{\Nn}}{\kappa_{\Nn}}
\simeq-1.08\times10^{-2}\,\mathrm{T^{-1}},
\]
\[
\frac{\zeta}{\kappa_{\Nn}}
\simeq
\frac{\Lambda_{\Nn}/\kappa_{\Nn}
      -\Delta\Nn/\Delta B_z}{\Delta\dR/\Delta B_z}
\simeq3.49\times10^{-3}\,\mathrm{nm^{-1}}.
\]
This estimate uses the specified field interval and the calibrated-state
coefficients above. The same field pair gives
\[
\frac{\Delta\Nn}{\Delta\dR}=-1.43893\times10^{-3}\,\mathrm{nm^{-1}},
\qquad \Qs(0)=0.16800.
\]
Section~\ref{sm:static-llg} compares this interval-based estimate with $\Nn$ measured
along the fixed-$q$ states and during the LLG oscillation.

\section{Radius difference under exchange modulation}
\label{sm:exchange-modulation}

We derive the local estimate used in main-text Eq.~(19). Let
$\mathbf m_0(\boldsymbol\xi)$ be a fixed circular N\'eel reference profile,
and set
\[
\begin{aligned}
\mathbf m_\eta(x,y)&=s_\eta\mathbf m_0(\boldsymbol\xi),\\
\boldsymbol\xi&=(x/b_\eta,y/a_\eta),\qquad s_A=+1,\ s_B=-1.
\end{aligned}
\]
This estimate retains the exchange, anisotropy, and DMI of one sublattice.
All energies below include the film thickness $t$ and are measured in
joules. Define
\[
\begin{aligned}
I_{\rm ex}&=\int d^2\xi\,|\partial_{\xi_x}\mathbf m_0|^2
            =\int d^2\xi\,|\partial_{\xi_y}\mathbf m_0|^2,\\
I_K&=\int d^2\xi\,(1-m_{0z}^2),\\
I_D&=-\int d^2\xi\,
 (m_{0z}\partial_{\xi_x}m_{0x}-m_{0x}\partial_{\xi_x}m_{0z}).
\end{aligned}
\]
The corresponding $y$ integral also equals $-I_D$.
The reference N\'eel orientation is chosen so that $DI_D>0$.
Using $dx\,dy=a_\eta b_\eta\,d^2\xi$ gives
\begin{equation}
\begin{aligned}
E_{{\rm ex},\eta}
 &=tI_{\rm ex}\left(A_{\eta x}\frac{a_\eta}{b_\eta}
                   +A_{\eta y}\frac{b_\eta}{a_\eta}\right),\\
E_{K,\eta}&=tKI_Ka_\eta b_\eta,\\
E_{D,\eta}&=-tDI_D(a_\eta+b_\eta).
\end{aligned}
\label{eq:local_energy_components_sm}
\end{equation}
Let $\tau_\eta=b_\eta/a_\eta$ and $R_\eta=\sqrt{a_\eta b_\eta}$.
Then
\begin{equation}
\begin{aligned}
E_\eta^{\rm loc}
&=t\left[I_{\rm ex}\left(\frac{A_{\eta x}}{\tau_\eta}
                    +A_{\eta y}\tau_\eta\right)\right.\\
&\qquad\left.+KI_KR_\eta^2-DI_DR_\eta g(\tau_\eta)\right],\\
g(\tau)&=\sqrt\tau+\frac1{\sqrt\tau}.
\end{aligned}
\label{eq:E_St}
\end{equation}
Minimizing only the exchange term at $\delta=0$ gives
\begin{equation}
\tau_A=\sqrt\rho,\qquad \tau_B=\rho^{-1/2},\qquad
E_{{\rm ex},\eta}^{\min}=2tI_{\rm ex}\bar A.
\label{eq:source_S7}
\end{equation}
At prescribed $\tau_\eta$, the radius minimizes
\begin{equation}
\begin{aligned}
\frac{\partial E_\eta^{\rm loc}}{\partial R_\eta}
 &=t[2KI_KR_\eta-DI_Dg(\tau_\eta)]=0,\\
R_\eta&=R_0g(\tau_\eta),\qquad
R_0=\frac{DI_D}{2KI_K}.
\end{aligned}
\label{eq:Rg_sm}
\end{equation}
The radius minimum exists for $K I_K>0$ and $DI_D>0$.
Since $g(\tau)=g(1/\tau)$, reciprocal aspect ratios give equal radii
within this trial family. The numerical radii instead minimize the full
energy, including intersublattice exchange and magnetostatics.

At $B_z=0$, apply the same relative modulation to both sublattices:
\begin{equation}
\begin{aligned}
(A_{Ax},A_{Ay})&=(A_1(1+\delta),A_2(1-\delta)),\\
(A_{Bx},A_{By})&=(A_2(1+\delta),A_1(1-\delta)).
\end{aligned}
\label{eq:source_S36}
\end{equation}
Thus $\Delta A_{\eta x}/A_{\eta x}^{(0)}=\delta$ and
$\Delta A_{\eta y}/A_{\eta y}^{(0)}=-\delta$.
The conversion from $\delta$ to lattice strain requires material-specific
exchange coefficients. The geometric means are
\[
\sqrt{A_{\eta x}A_{\eta y}}
=\bar A\sqrt{1-\delta^2}
=\bar A[1-\delta^2/2+O(\delta^4)].
\]
The exchange-only aspect-ratio minima become
\begin{equation}
\begin{aligned}
\tau_A(\delta)
 &=\sqrt\rho\sqrt{\frac{1+\delta}{1-\delta}}
 =\sqrt\rho(1+\delta)+O(\delta^2),\\
\tau_B(\delta)
 &=\rho^{-1/2}\sqrt{\frac{1+\delta}{1-\delta}}
 =\rho^{-1/2}(1+\delta)+O(\delta^2).
\end{aligned}
\label{eq:source_S37}
\end{equation}
Inserting these into the local radius estimate gives
\begin{equation}
\begin{aligned}
\dR&=R_0[g(\tau_A)-g(\tau_B)]\\
&=R_0(\rho^{1/4}-\rho^{-1/4})
\left[\left(\frac{1+\delta}{1-\delta}\right)^{1/4}
-\left(\frac{1-\delta}{1+\delta}\right)^{1/4}\right].
\end{aligned}
\label{eq:source_S38}
\end{equation}
For small $\delta$,
\begin{equation}
\dR\simeq R_0(\rho^{1/4}-\rho^{-1/4})\delta.
\label{eq:strain_sm}
\end{equation}
The separate radius derivatives in this approximation are
\[
\left.\frac{dR_A}{d\delta}\right|_0
=-\left.\frac{dR_B}{d\delta}\right|_0
=\frac{R_0}{2}(\rho^{1/4}-\rho^{-1/4}).
\]
They have opposite signs for $\rho>1$ and vanish at $\rho=1$.

The full-energy calculations use
\[
\delta\in\{0,\pm0.02,\pm0.04,\pm0.06\},\qquad B_z=0,
\]
with $K,D$ and the other unmodulated parameters fixed. Near $\delta=0$,
\[
\dR(\delta)\simeq c_\epsilon+\chi_\epsilon\delta,
\qquad
\chi_\epsilon=\left.\frac{d\dR}{d\delta}\right|_0,
\]
where the numerical slope is estimated by fitting the computed states
with $|\delta|\le0.02$. The fitted values are
\[
\begin{array}{c|r}
\rho&\chi_\epsilon\ (\mathrm{nm})\\ \hline
1&1.30\times10^{-11}\\
1.5&3.15\times10^{-1}\\
2&3.85\times10^{-1}\\
2.5&6.48\times10^{-1}\\
3&8.29\times10^{-1}\\
3.5&1.09\\
4&1.19
\end{array}
\]
At $\rho=1$, symmetry gives $\chi_\epsilon=0$.
The numerical fit gives
$c_\epsilon=3.82\times10^{-8}\,\mathrm{nm}$ and the residual slope
shown above.

\section{Local intersublattice energy of separated contours}
\label{sm:wall-energy}

The excess intersublattice energy is
\[
E_{\rm inter}=A_0t\int d^2r\,[1+\mathbf m_A\cdot\mathbf m_B].
\]
For a local estimate, neglect the wall curvature across each cross-section,
replace the normal separation by $u(\varphi)$, and use the line element
$ds\simeq\Rbar\,d\varphi$. The two walls have a common width $\lambda$.
Then
\begin{equation}
E_{\rm inter}\simeq A_0t\lambda\int_0^{2\pi}
\Rbar\,F\!\left(\frac{u(\varphi)}{\lambda}\right)d\varphi.
\label{eq:Einter}
\end{equation}
Let $\ell$ be the normal coordinate and define
\[
\begin{aligned}
\mathbf n_0(\xi)&=(\operatorname{sech}\xi,0,\tanh\xi),\\
\mathbf m_A(\ell)&=\mathbf n_0\!\left(\frac{\ell-u/2}{\lambda}\right),\\
\mathbf m_B(\ell)&=-\mathbf n_0\!\left(\frac{\ell+u/2}{\lambda}\right).
\end{aligned}
\]
With $\xi=\ell/\lambda$ and $v=u/\lambda$,
\begin{equation}
\begin{aligned}
F(v)
&=\int_{-\infty}^{\infty}d\xi\,
 [1-\mathbf n_0(\xi-v/2)\cdot\mathbf n_0(\xi+v/2)]\\
&=(\cosh v-1)\int_{-\infty}^{\infty}
\frac{d\xi}{\cosh(\xi-v/2)\cosh(\xi+v/2)}\\
&=2v\tanh(v/2)\\
&=v^2-\frac{v^4}{12}+O(v^6),\qquad
F(v)\sim2|v|\quad(|v|\gg1).
\end{aligned}
\label{eq:Ftanh}
\end{equation}

At $\dR=0$ in AM, use
$v(\varphi)=2\Rbar\Qs\cos2\varphi/\lambda$.
At fixed $\Qs$ and $\lambda$,
\[
\frac{\partial E_{\rm inter}}{\partial\Rbar}
\simeq A_0t\lambda\int_0^{2\pi}[F(v)+vF'(v)]\,d\varphi,
\]
\[
F'(v)=2\tanh(v/2)+v\operatorname{sech}^2(v/2).
\]
For the antiferromagnetic coupling $A_0>0$,
$F(v)\ge0$ and $vF'(v)\ge0$.
Thus the local estimate increases with $\Rbar$ when $\Qs\ne0$.
In the zero-field PAR state,
$u=0$ and $F(0)=0$, so $E_{\rm inter}=0$.

For an in-plane unit normal $\hat{\mathbf n}=(n_x,n_y)$,
the single-sublattice wall-width estimates are
\[
\begin{aligned}
\lambda_A(\hat{\mathbf n})
 &=\sqrt{\frac{A_1n_x^2+A_2n_y^2}{K}},\\
\lambda_B(\hat{\mathbf n})
 &=\sqrt{\frac{A_2n_x^2+A_1n_y^2}{K}}.
\end{aligned}
\]
Along $x$,
\[
\begin{aligned}
\lambda_A&=\lambda\rho^{1/4},\qquad
\lambda_B=\lambda\rho^{-1/4},\\
\frac{\lambda_A}{\lambda_B}&=\sqrt\rho=2\quad(\rho=4).
\end{aligned}
\]
The corresponding estimates are equal between sublattices in PAR.
The common-$\lambda$ calculation above therefore omits the difference
between the two directional wall-width scales. The numerical states
minimize the full energy without imposing a common wall width.

\section{Energy of states minimized at fixed $q$}
\label{sm:constrained-energy}

For AM at $\rho=2$ and $B_z=0$, the constrained coordinate is
\begin{equation}
\begin{aligned}
q&=\sqrt{S_A/\pi}-\sqrt{S_B/\pi},\\
S_\eta&=\frac{h^2}{2}\sum_{ij}(1+s_\eta m_{\eta z,ij}),\\
s_A&=+1,\qquad s_B=-1.
\end{aligned}
\label{eq:source_S71}
\end{equation}
For prescribed $q^*$, we minimize
\[
E_{\rm aug}
=E+\Lambda_q(q-q^*)+\frac{k_p}{2}(q-q^*)^2,
\]
where $[\Lambda_q]=\mathrm{J/m}$ and $[k_p]=\mathrm{J/m^2}$.
The constraint gradient is obtained from
\[
\frac{\partial q}{\partial\mathbf m_{\eta,ij}}
=\frac{h^2}{4\sqrt{\pi S_\eta}}\hat{\mathbf e}_z,
\qquad \eta=A,B,
\]
\[
\mathbf g_{\eta,ij}^{\rm aug}
=-\frac{1}{h^2t}\left[
\frac{\partial E}{\partial\mathbf m_{\eta,ij}}
+[\Lambda_q+k_p(q-q^*)]\frac{\partial q}{\partial\mathbf m_{\eta,ij}}
\right].
\]
The tangent projection and norms are defined as in Sec.~\ref{sm:static-methods}, now using
$\mathbf g^{\rm aug}$. Accepted states satisfy
\[
\begin{aligned}
g_{\max}^{\rm aug}&\le1\,\mathrm{J/m^3},\\
g_{\rm rms}^{\rm aug}&\le0.1\,\mathrm{J/m^3},\\
|q-q^*|&\le0.002\,\mathrm{nm}.
\end{aligned}
\]
When the constraint tolerance is not met, the multiplier is updated by
\[
\Lambda_q\leftarrow\Lambda_q+k_p(q-q^*).
\]
From the fourth outer iteration onward, the update also includes
$k_p\leftarrow1.5k_p$.
After minimization, we record the physical energy $E$ and measure
$\dR=\sqrt{a_Ab_A}-\sqrt{a_Bb_B}$ from the contours. The constraint
coordinate $q$ includes the wall magnetization through $S_\eta$ and is
not set equal to $\dR$.

The accepted targets are
\[
q_j^*=0.25j\,\mathrm{nm},\qquad j=-12,\ldots,12.
\]
These 25 states retain both cores and give $|\dR|\lesssim2.8046\,\mathrm{nm}$.
At $q^*=\pm3.25\,\mathrm{nm}$ one core disappears.
For the states with $|\dR_j|\le0.8\,\mathrm{nm}$, the least-squares fit is
\[
\begin{aligned}
&(E_c,b_R,\kappa_{\rm con})\\
&\quad=\underset{c,b,k}{\operatorname{argmin}}
\sum_j\left[E_j-c-b\dR_j-\frac{k}{2}(\dR_j)^2\right]^2.
\end{aligned}
\]
It gives
\[
\kappa_{\rm con}=12.1658\,\mathrm{mJ/m^2}.
\]
The intercept $E_c$ and linear coefficient $b_R$ are fitted independently.
The reported curvature is twice the quadratic polynomial coefficient. This coefficient describes
the energy along states minimized at fixed $q$, plotted against their
measured $\dR$. Section~\ref{sm:static-llg} obtains $P$ from the same states.

\section{LLG integration, helicity, and oscillation fits}
\label{sm:llg}

Let $\mathbf r_{\eta,j}^{(0)}$ be the 181 sampling positions on the
zero-field contour,
\[
\mathbf r_{\eta,j}^{(0)}
=r_\eta^{(0)}(\varphi_j)(\cos\varphi_j,\sin\varphi_j).
\]
The positions remain fixed during the LLG calculation. The spin components
at these positions are interpolated bilinearly. Define
\[
Z_\eta(t)=\sum_j
\left[m_{\eta x}(\mathbf r_{\eta,j}^{(0)},t)
+\mathrm i m_{\eta y}(\mathbf r_{\eta,j}^{(0)},t)\right]
 e^{-\mathrm i\varphi_j}.
\]
This is identical to the stated $|\mathbf m_{\eta\parallel}|$-weighted
sum because
\[
|\mathbf m_{\eta\parallel}|
 e^{\mathrm i\operatorname{atan2}(m_{\eta y},m_{\eta x})}
=m_{\eta x}+\mathrm i m_{\eta y}.
\]
The helicity deviation is
\[
\phi_\eta(t)
=\operatorname{unwrap}[\arg Z_\eta(t)]-\phi_\eta^{(0)},
\qquad
\phi_\eta^{(0)}=\arg Z_\eta^{(0)},
\]
where $Z_\eta^{(0)}$ is evaluated in the zero-field equilibrium state. 

The main-text angles are
\[
\bar\phi(t)=\frac{\phi_A(t)+\phi_B(t)}2,
\qquad
\phi_-(t)=\frac{\phi_A(t)-\phi_B(t)}2.
\]
For a common spin rotation by $\beta$,
\[
Z_\eta\mapsto e^{\mathrm i\beta}Z_\eta,
\qquad \phi_\eta\mapsto\phi_\eta+\beta,
\qquad \bar\phi\mapsto\bar\phi+\beta.
\]

The pulse ends at $t=0$:
\[
B_z(t)=
\begin{cases}
0.05\,\mathrm T,&-10\,\mathrm{ps}\le t<0,\\
0,&0\le t\le2\,\mathrm{ns}.
\end{cases}
\]
We integrate main-text Eq.~(3) with $\alpha=10^{-3}$ and
$\Delta t_{\rm LLG}=4\,\mathrm{fs}$.
Writing its right-hand side for the full spin array as
$\mathbf v[\mathbf m,t]$, the fourth-order Runge--Kutta step is
\[
\begin{aligned}
\mathbf k_1&=\mathbf v[\mathbf m^\ell,t_\ell^{\rm LLG}],\\
\mathbf k_2&=\mathbf v[\mathbf m^\ell+\tfrac12\Delta t_{\rm LLG}\mathbf k_1,
 t_\ell^{\rm LLG}+\tfrac12\Delta t_{\rm LLG}],\\
\mathbf k_3&=\mathbf v[\mathbf m^\ell+\tfrac12\Delta t_{\rm LLG}\mathbf k_2,
 t_\ell^{\rm LLG}+\tfrac12\Delta t_{\rm LLG}],\\
\mathbf k_4&=\mathbf v[\mathbf m^\ell+\Delta t_{\rm LLG}\mathbf k_3,
 t_\ell^{\rm LLG}+\Delta t_{\rm LLG}],\\
\mathbf m^{\ell+1}_{\rm pre}
 &=\mathbf m^\ell+\frac{\Delta t_{\rm LLG}}6
  (\mathbf k_1+2\mathbf k_2+2\mathbf k_3+\mathbf k_4).
\end{aligned}
\]
The complete effective field, including magnetostatics, is recomputed
in every evaluation of $\mathbf v$. Each sublattice spin is then
normalized separately:
\[
\mathbf m_{\eta,ij}^{\ell+1}
=\frac{\mathbf m_{\eta,ij,\rm pre}^{\ell+1}}
{|\mathbf m_{\eta,ij,\rm pre}^{\ell+1}|}.
\]
The seven values of $\rho$ in both AM and PAR give 14 LLG calculations
in the $96\,\mathrm{nm}$ periodic cell.
The saved series contains
\[
\begin{aligned}
N_t&=4001,\qquad \Delta t_{\rm save}=0.5\,\mathrm{ps},\\
t_n&=n\Delta t_{\rm save},\qquad n=0,\ldots,N_t-1.
\end{aligned}
\]
For a coordinate $X$ to which the damped-cosine fit is applied, the model is
\[
\begin{aligned}
X_{\rm fit}(t)&=c_X+A_Xe^{-t/\tau_X}\cos(2\pi f_Xt+\psi_X),\\
&\hspace{2em}0\le t\le2\,\mathrm{ns}.
\end{aligned}
\]
Here $|A_X|$ is the envelope amplitude at $t=0$, $\tau_X$ is the decay
time, $f_X$ is the frequency, $\psi_X$ is the phase at $t=0$, and $c_X$
is a constant offset. The Fourier frequencies and amplitudes are defined
below.

For AM at $\rho=2$, the time-domain fits give
\[
f_{\dR}=49.52188\,\mathrm{GHz},\qquad
\psi_{\bar\phi}-\psi_{\dR}=90.01048^\circ.
\]
For each sampled coordinate $X_n=X(t_n)$, define
\[
\langle X\rangle_t=\frac1{N_t}\sum_{n=0}^{N_t-1}X_n,
\qquad f_k=\frac{k}{N_t\Delta t_{\rm save}},
\]
\[
\begin{aligned}
\widehat X(f_k)&=
\frac{2}{\sum_{n=0}^{N_t-1}w_n}
\sum_{n=0}^{N_t-1}w_n[X_n-\langle X\rangle_t]
 e^{-2\pi\mathrm i kn/N_t},\\
A_X^{\rm F}(f_k)&=|\widehat X(f_k)|.
\end{aligned}
\]
The symmetric Hann weights are
\[
w_n=\frac12\left[1-\cos\left(\frac{2\pi n}{N_t-1}\right)\right],
\qquad n=0,\ldots,N_t-1.
\]
The transform uses $N_t$ points without zero padding.
The initial peak search includes every nonzero positive frequency:
\[
k_{\rm init}=\underset{k\in\mathcal K}{\operatorname{argmax}}
A_{\dR}^{\rm F}(f_k),\qquad
\mathcal K=\{1,\ldots,(N_t-1)/2\}.
\]
After the time-domain fit, reported Fourier amplitudes use
\[
k_* = \underset{k\in\mathcal K}{\operatorname{argmin}}
|f_k-f_{\dR}|.
\]
For the $\rho=2$ and $4$ AM data reported here, $k_*=k_{\rm init}$.
 The positive-frequency spacing is
\[
\Delta f=\frac1{N_t\Delta t_{\rm save}}
=0.499875\ldots\,\mathrm{GHz}.
\]
For the selected Fourier frequency at $\rho=2$,
\[
k_*=99,\qquad f_{k_*}=49.4876\,\mathrm{GHz}.
\]
The amplitudes evaluated at this same frequency are
\[
\begin{array}{c|r|c}
X&A_X^{\rm F}(f_{k_*})&\text{unit}\\ \hline
\dR&0.2068795&\mathrm{nm}\\
\bar\phi&0.03257398&\mathrm{rad}\\
\phi_-&5.956134\times10^{-8}&\mathrm{rad}\\
\Nn&1.445541\times10^{-5}&1\\
\Rbar&2.354173\times10^{-7}&\mathrm{nm}
\end{array}
\]
Therefore,
\[
\begin{aligned}
\frac{A_{\phi_-}^{\rm F}(f_{k_*})}{A_{\bar\phi}^{\rm F}(f_{k_*})}
 &=1.82849\times10^{-6},\\
\frac{A_{\Rbar}^{\rm F}(f_{k_*})}{A_{\dR}^{\rm F}(f_{k_*})}
 &=1.13794\times10^{-6},\\
\frac{A_{\Nn}^{\rm F}(f_{k_*})}{A_{\dR}^{\rm F}(f_{k_*})}
 &=6.98736\times10^{-5}\,\mathrm{nm^{-1}}.
\end{aligned}
\]
These comparisons concern $f_{k_*}$.

The projections of $\bar\phi$ and $\Nn$ fix the frequency and decay time to those obtained
from $\dR(t)$, with a free offset and cosine and sine coefficients
for each coordinate. The phase of $\dR$ is represented with a positive
amplitude. For the $\Nn$ projection, define
\[
\begin{aligned}
\theta_{\rm osc}(t)&=2\pi f_{\dR}t+\psi_{\dR},\\
C(t)&=e^{-t/\tau_{\dR}}\cos\theta_{\rm osc}(t),\\
S(t)&=e^{-t/\tau_{\dR}}\sin\theta_{\rm osc}(t).
\end{aligned}
\]
The projected series is
\[
\Nn^{\rm proj}(t)=c_e+a_e C(t)+b_e S(t).
\]
The unweighted least-squares projection includes a free constant:
\[
(c_e,a_e,b_e)=\underset{c,a,b}{\operatorname{argmin}}
\sum_n[\Nn(t_n)-c-aC(t_n)-bS(t_n)]^2.
\]
Its amplitude and phase relative to $\dR$ are
\[
A_e^{\rm proj}=\sqrt{a_e^2+b_e^2},\qquad
\Delta\psi_e^{\rm proj}=\operatorname{atan2}(-b_e,a_e).
\]
The reported values are
\[
\begin{aligned}
\frac{A_e^{\rm proj}}{|A_{\dR}|}
&=1.236\times10^{-4}\,\mathrm{nm^{-1}},\\
\Delta\psi_e^{\rm proj}&=-0.653^\circ,
\qquad \tau_{\dR}=3.2165\,\mathrm{ns}.
\end{aligned}
\]
The quoted decay time is inherited from the $\dR$ fit.

For each value of $\rho$, let $k_*$ denote the frequency index selected
as above. The surrounding $\phi_-$ amplitude is defined by
\[
\begin{aligned}
A_{\phi_-,\rm bg}^{\rm F}
&=\operatorname{median}_{j\in\mathcal J_{\rm bg}}
 A_{\phi_-}^{\rm F}(f_{k_*+j}),\\
\mathcal J_{\rm bg}&=\{-5,-4,-3,-2,2,3,4,5\}.
\end{aligned}
\]
For AM at $\rho=4$,
\[
\frac{A_{\phi_-}^{\rm F}(f_{k_*})}{A_{\bar\phi}^{\rm F}(f_{k_*})}
=1.53392\times10^{-5},\qquad
\frac{A_{\phi_-}^{\rm F}(f_{k_*})}{A_{\phi_-,\rm bg}^{\rm F}}=23.84.
\]
Section~\ref{sm:static-llg} compares the oscillation with the prediction from static
coefficients.

\section{Berry coupling of radius difference and helicity}
\label{sm:berry}

Let $\mathbf m_\eta^{(q)}(\mathbf r)$ be the zero-field states minimized
at prescribed $q$ in Sec.~\ref{sm:constrained-energy}. Locally near $q=0$, the trial motion is
\[
\begin{aligned}
\mathbf m_\eta(\mathbf r,t')
&=R_z[\bar\phi(t')]\mathbf m_\eta^{(q(t'))}(\mathbf r),\\
\dR(0)&=0,\qquad \left.\frac{d\dR}{dq}\right|_0\ne0.
\end{aligned}
\]
Here $R_z$ rotates the spins, and $t'$ is
the time variable. 
The prefactor $M_st$ below contains the film thickness.
We assume a smooth local continuation of this family, so either $q$ or
its measured $\dR$ can parametrize the radial motion.

Write the local unit spin as
\[
\mathbf m_\eta=(\sin\theta_\eta\cos\Phi_\eta,
\sin\theta_\eta\sin\Phi_\eta,\cos\theta_\eta).
\]
The Berry action is
\begin{equation}
S_B=\frac{M_st}{\gamma}\sum_{\eta=A,B}
\int dt'\int d^2r\,
\Omega_\eta(\theta_\eta)\dot\Phi_\eta.
\label{eq:source_S45}
\end{equation}
The overdot denotes differentiation with respect to time.
We choose
\begin{equation}
\begin{aligned}
\Omega_\eta&=-\cos\theta_\eta-s_\eta,\\
\Omega_A&=-(1+\cos\theta_A),\qquad
\Omega_B=1-\cos\theta_B.
\end{aligned}
\label{eq:source_S46}
\end{equation}
Then $\partial_{\theta_\eta}\Omega_\eta=\sin\theta_\eta$, with the
precession sign of main-text Eq.~(3), and
$\Omega_\eta\to0$ in the backgrounds $\cos\theta_\eta\to-s_\eta$.
Define
\begin{equation}
I_A=\int d^2r\,(1+\cos\theta_A),\qquad
I_B=\int d^2r\,(1-\cos\theta_B).
\label{eq:source_S47}
\end{equation}
These integrands vanish in the corresponding backgrounds and satisfy
$I_\eta=2S_\eta$.

For uniform spin-angle increments,
\[
\Phi_\eta(\mathbf r,t')
=\Phi_\eta^{(q(t'))}(\mathbf r)+\phi_\eta(t'),\qquad
\dot\Phi_\eta
=\partial_q\Phi_\eta^{(q)}\dot q+\dot\phi_\eta.
\]
The term proportional to $\dot q$ is $\mathcal A(q)\dot q$, where
\[
\mathcal A(q)=\frac{M_st}{\gamma}\sum_\eta\int d^2r\,
\Omega_\eta(\theta_\eta^{(q)})\partial_q\Phi_\eta^{(q)}.
\]
Locally along this one-parameter family,
\[
\mathcal A(q)\dot q
=\frac{d}{dt'}\left[\int_0^{q(t')}\mathcal A(q')\,dq'\right].
\]
After removing this total time derivative, the Berry action is
\begin{equation}
S_B=\frac{M_st}{\gamma}\int dt'
[-I_A\dot\phi_A+I_B\dot\phi_B].
\label{eq:SB_sm}
\end{equation}
Using $\phi_A=\bar\phi+\phi_-$ and
$\phi_B=\bar\phi-\phi_-$ gives
\begin{equation}
S_B=-\frac{M_st}{\gamma}\int dt'
[(I_A-I_B)\dot{\bar\phi}+(I_A+I_B)\dot\phi_-].
\label{eq:SB2_sm}
\end{equation}
For these uniform increments about the zero-field helicities,
\[
\mathcal P:(\phi_A,\phi_B)\mapsto(-\phi_B,-\phi_A),\qquad
(\bar\phi,\phi_-)\mapsto(-\bar\phi,\phi_-).
\]
The exact identity
\[
M_st(I_A-I_B)
=M_st\int d^2r\,(m_{Az}+m_{Bz})=\mu_z
\]
connects the common spin rotation to the net moment.
For the locally $\mathcal P$-related states at $q$ and $-q$,
$\mu_z$ and $\dR$ are odd. Thus
\begin{equation}
I_A-I_B
=\left.\frac{d(I_A-I_B)}{d\dR}\right|_0\dR+O(\dR^3).
\label{eq:source_S50}
\end{equation}
The common-rotation trial has $\phi_-=0$. Its leading Berry term is
\[
L_B=-P\dR\dot{\bar\phi},\qquad
P=\frac{M_st}{\gamma}\left.\frac{d(I_A-I_B)}{d\dR}\right|_0
=\frac1\gamma\left.\frac{d\mu_z}{d\dR}\right|_0.
\]
Both derivatives are taken along the same fixed-$q$ family. Equivalently,
\[
P=\left.\frac1\gamma
\frac{d\mu_z/dq}{d\dR/dq}\right|_{q=0}.
\]
The quadratic term in the radial energy is
$\tfrac12\kappa_{\rm con}\dR(q)^2$, with $\kappa_{\rm con}$ obtained
from the polynomial fit in Sec.~\ref{sm:constrained-energy}.
The common spin rotation leaves $m_{\eta z}$ and therefore the measured
$\dR,\Rbar,\Qs,\Nn$ unchanged. In particular,
$\Nn(t')=\Nn[\mathbf m^{(q(t'))}]$ within this trial family.
The radial energy and Berry term enter the Lagrangian in Sec.~\ref{sm:two-coordinate}.

\section{Energy of a common spin rotation}
\label{sm:rotation-energy}

For the zero-field state at $\rho=2$, rotate each spin by the same angle
while keeping its position fixed:
\[
\mathbf m_\eta(\mathbf r;\bar\phi)
=R_z(\bar\phi)\mathbf m_\eta^{(0)}(\mathbf r),
\]
\[
R_z(\bar\phi)=
\begin{pmatrix}
\cos\bar\phi&-\sin\bar\phi&0\\
\sin\bar\phi&\cos\bar\phi&0\\
0&0&1
\end{pmatrix}.
\]
The exchange, anisotropy, and intersublattice terms are unchanged.
The separate second derivatives are
\[
\begin{aligned}
\kappa_{\bar\phi}^{\rm DMI}
 &=\left.\frac{d^2E_D[R_z(\bar\phi)\mathbf m^{(0)}]}{d\bar\phi^2}\right|_0,\\
\kappa_{\bar\phi}^{\rm ms}
 &=\left.\frac{d^2E_{\rm ms}[R_z(\bar\phi)\mathbf m^{(0)}]}{d\bar\phi^2}\right|_0.
\end{aligned}
\]
The $\mathcal Q$ symmetry in Sec.~\ref{sm:two-coordinate} makes the energy even in
$\bar\phi$, so
\[
E(\bar\phi)-E(0)=\frac12\kappa_{\bar\phi}\bar\phi^2+O(\bar\phi^4).
\]
Their sum and the circular-wall estimate are
\begin{equation}
\kappa_{\bar\phi}
=\kappa_{\bar\phi}^{\rm DMI}+\kappa_{\bar\phi}^{\rm ms},
\qquad
\kappa_{\bar\phi}^{\rm DMI}\simeq4\pi^2D\Rbar t.
\label{eq:source_S51}
\end{equation}
The final estimate is derived below for circular walls with
$\lambda/R_\eta\ll1$.
The numerical coefficient is obtained by rotating
the complete calculated state.

For a circular N\'eel texture, write
\[
\Phi_\eta=\varphi+\phi_\eta^{(0)}+\bar\phi,
\qquad
\int_0^\infty\theta_\eta'(r)\,dr=s_\eta\pi,
\]
with core and background polarities $m_{\eta z}(0)=s_\eta$ and
$m_{\eta z}(\infty)=-s_\eta$.
For the chosen $D>0$, the lower-energy N\'eel orientations satisfy
$\cos\phi_\eta^{(0)}=-s_\eta$.
The DMI energy of each sublattice is
\[
\begin{aligned}
E_{D,\eta}(\bar\phi)
&=2\pi Dt\cos(\phi_\eta^{(0)}+\bar\phi)\\
&\quad\times\int_0^\infty[r\theta_\eta'(r)
              +\sin\theta_\eta\cos\theta_\eta]dr.
\end{aligned}
\]
For a smooth wall of width $\lambda\ll R_\eta$,
\[
\begin{aligned}
\int_0^\infty r\theta_\eta'(r)dr&\simeq s_\eta\pi R_\eta,\\
\left|\int_0^\infty\sin\theta_\eta\cos\theta_\eta\,dr\right|
&=O(\lambda).
\end{aligned}
\]
Keeping the leading term gives
\[
\begin{aligned}
E_{D,\eta}(\bar\phi)&\simeq-2\pi^2DtR_\eta\cos\bar\phi,\\
E_D(\bar\phi)&\simeq-4\pi^2Dt\Rbar\cos\bar\phi,\\
\left.\frac{d^2E_D}{d\bar\phi^2}\right|_0&\simeq4\pi^2Dt\Rbar.
\end{aligned}
\]

The numerical rotations use
\[
\bar\phi\in\{0,\pm0.01,\pm0.02,\pm0.04,\pm0.08\}\,\mathrm{rad}.
\]
For each contribution $a\in\{D,{\rm ms}\}$, the unweighted fit is
\[
\begin{aligned}
&(E_{a,\rm off},b_a,k_a)=\underset{c,b,k}{\operatorname{argmin}}\\
&\quad\sum_{|\bar\phi_j|\le0.04}
\left[E_a(\bar\phi_j)-c-b\bar\phi_j-\frac{k}{2}\bar\phi_j^2\right]^2.
\end{aligned}
\]

The coefficients $k_D$ and $k_{\rm ms}$ estimate
$\kappa_{\bar\phi}^{\rm DMI}$ and $\kappa_{\bar\phi}^{\rm ms}$.

Quadratic fits using $|\bar\phi|\le0.04\,\mathrm{rad}$ give
\begin{equation}
\begin{aligned}
\kappa_{\bar\phi}^{\rm DMI}&=4.09917880\times10^{-19}\,\mathrm J,\\
\kappa_{\bar\phi}^{\rm ms}&=-1.43448804\times10^{-21}\,\mathrm J,\\
\kappa_{\bar\phi}&=4.08483392\times10^{-19}\,\mathrm J.
\end{aligned}
\label{eq:source_S52}
\end{equation}
Varying the maximum included angle from $0.02$ to $0.08\,\mathrm{rad}$
gives the range
\[
\kappa_{\bar\phi}\in[4.08317,4.08526]\times10^{-19}\,\mathrm J,
\]
\[
\frac{4.08526-4.08317}{4.08483392}\simeq0.051\%.
\]

For a magnetization source $\mathbf U$, denote the quadratic magnetostatic
energy of Sec.~\ref{sm:magnetostatics} by $E_{\rm ms}[\mathbf U]$. Under the common rotation,
\[
\begin{aligned}
\mathbf M_\parallel(\bar\phi)
&=\cos\bar\phi\,\mathbf M_\parallel^{(0)}\\
&\quad+\sin\bar\phi\,(\hat{\mathbf e}_z\times\mathbf M_\parallel^{(0)}),\\
M_z(\bar\phi)&=M_z^{(0)}.
\end{aligned}
\]
Using the quadratic energy and $N_{xz}=N_{yz}=0$ gives
\[
\kappa_{\bar\phi}^{\rm ms}
=2\left[E_{\rm ms}[\hat{\mathbf e}_z\times\mathbf M_\parallel^{(0)}]
        -E_{\rm ms}[\mathbf M_\parallel^{(0)}]\right].
\]
Each energy on the right is nonnegative, but their difference can have
either sign. For the calculated state,
\[
\frac{\kappa_{\bar\phi}^{\rm ms}}
{\kappa_{\bar\phi}^{\rm DMI}}=-0.350\%,
\qquad \kappa_{\bar\phi}>0.
\]

\section{Symmetry and equations of the two-coordinate model}
\label{sm:two-coordinate}

Under $\mathcal P$, $(\dR,\bar\phi)\mapsto(-\dR,-\bar\phi)$, so
$\dR\bar\phi$ is unchanged. To test this potential-energy term, use the
static transformation
\begin{equation}
\mathcal Q:\quad
\mathbf m_\eta(x,y)\mapsto
(-m_{\eta x},m_{\eta y},m_{\eta z})(-x,y).
\label{eq:source_S56}
\end{equation}
This is reflection in $x$ combined with time reversal. Let
$R_x=\operatorname{diag}(-1,1,1)$.
The local energy densities satisfy
\[
\begin{aligned}
w_a[\mathcal Q\mathbf m](x,y)&=w_a[\mathbf m](-x,y),\\
a&\in\{{\rm ex},K,D,{\rm inter},Z\}.
\end{aligned}
\]
with $B_z$ unchanged. For magnetostatics,
\[
\widehat{\mathbf M}'(\mathbf k)=R_x\widehat{\mathbf M}(R_x\mathbf k),
\qquad N(R_x\mathbf k)=R_xN(\mathbf k)R_x^T,
\]
so $E_{\rm ms}[\mathcal Q\mathbf m]=E_{\rm ms}[\mathbf m]$.
The contour and spin angles transform as
\[
\begin{aligned}
r_\eta(\varphi)&\mapsto r_\eta(\pi-\varphi),\\
(\Rbar,\dR,\Qs,\Nn)&\mapsto(\Rbar,\dR,\Qs,\Nn),\\
(\bar\phi,\phi_-)&\mapsto(-\bar\phi,-\phi_-).
\end{aligned}
\]
For the selected zero-field family preserving $\mathcal Q$,
\[
E(\dR,\bar\phi)=E(\dR,-\bar\phi),\qquad
\left.\frac{\partial^2E}{\partial\dR\,\partial\bar\phi}\right|_0=0.
\]
Thus the quadratic potential contains no $\dR\bar\phi$ term.

Without damping, the quadratic Lagrangian for this trial family is
\begin{equation}
L_{\rm var}=-P\dR\dot{\bar\phi}
-\frac12\kappa_{\rm con}\dR^2
-\frac12\kappa_{\bar\phi}\bar\phi^2.
\label{eq:source_S54}
\end{equation}
The Euler--Lagrange equations give
\[
P\dot{\dR}=\kappa_{\bar\phi}\bar\phi,
\qquad
P\dot{\bar\phi}=-\kappa_{\rm con}\dR.
\]
Eliminating $\bar\phi$ gives
\begin{equation}
\frac{P^2}{\kappa_{\bar\phi}}\ddot{\dR}
=-\kappa_{\rm con}\dR,
\qquad m^*=\frac{P^2}{\kappa_{\bar\phi}}.
\label{eq:mstar_sm}
\end{equation}
Here $m^*$ is the mass associated with $\dR$, with units
$\mathrm{J\,s^2/m^2}=\mathrm{kg}$.
For $\kappa_{\rm con}>0$, $\kappa_{\bar\phi}>0$, and $P\ne0$,
\[
\omega_0=\frac{\sqrt{\kappa_{\rm con}\kappa_{\bar\phi}}}{|P|},
\qquad f_0=\frac{\omega_0}{2\pi}.
\]
Choosing $\dR(t')=A_{\dR}\cos(\omega_0t'+\psi_{\dR})$ gives
\[
\bar\phi(t')
=-\frac{P\omega_0}{\kappa_{\bar\phi}}A_{\dR}
\sin(\omega_0t'+\psi_{\dR}).
\]
Therefore the undamped model predicts
\[
\frac{|\bar\phi|}{|\dR|}
=\sqrt{\frac{\kappa_{\rm con}}{\kappa_{\bar\phi}}},
\qquad
\psi_{\bar\phi}-\psi_{\dR}=\operatorname{sgn}(P)\frac\pi2.
\]

Internal spin deformations can give skyrmion inertia \cite{SM-PhysRevLett.109.217201,SM-PhysRevB.90.174434}.
 Combining the piecewise $m_z$ integrals of Sec.~\ref{sm:moments} with the
circular-wall DMI estimate of Sec.~\ref{sm:rotation-energy} gives
\[
I_\eta\simeq2\pi R_\eta^2,\qquad
I_A-I_B\simeq4\pi\Rbar\dR,
\qquad
P\simeq\frac{4\pi M_st\Rbar}{\gamma}.
\]
If $\lambda/\Rbar\ll1$ and the magnetostatic rotation term is omitted,
\[
\kappa_{\bar\phi}\simeq4\pi^2Dt\Rbar,
\qquad
m^*\simeq\frac{4M_s^2t\Rbar}{\gamma^2D}.
\]
For the calculated reference state,
$\lambda/\Rbar\simeq4.85/3.87\simeq1.25$.
The numerical prediction in Sec.~\ref{sm:static-llg} uses $P$ and
$\kappa_{\bar\phi}$ evaluated from the complete spin configurations.

\section{Comparison with the LLG oscillation}
\label{sm:static-llg}

For AM at $\rho=2$, use the seven fixed-$q$ states with measured
$|\dR_j|\le0.8\,\mathrm{nm}$, selected as $|q^*|$ increases from zero.
The unweighted fit is
\[
(I_{\rm off},s_I)
=\underset{c,s}{\operatorname{argmin}}
\sum_{j=1}^{7}
[(I_{A,j}-I_{B,j})-c-s\dR_j]^2.
\]
The intercept $I_{\rm off}$ is independent. The slope $s_I$ has units
of length and estimates the derivative at zero:
\begin{equation}
\begin{aligned}
\left.\frac{d(I_A-I_B)}{d\dR}\right|_0
&\simeq s_I=6.42339469\times10^{-8}\,\mathrm m,\\
P&\simeq\frac{M_st}{\gamma}s_I
=2.11680052\times10^{-22}\,\mathrm{J\,s/m}.
\end{aligned}
\label{eq:source_S57}
\end{equation}

The static inputs at $\rho=2$ are
\[
\begin{aligned}
\kappa_{\rm con}&=0.01216576558\,\mathrm{J/m^2},\\
\kappa_{\bar\phi}&=4.08483392\times10^{-19}\,\mathrm J.
\end{aligned}
\]
together with $P$ above. The first two are obtained from the fixed-$q$
energy fit and the common spin rotation, respectively. The two-coordinate
model then gives
\begin{equation}
\begin{aligned}
f_0&=\frac{\sqrt{\kappa_{\rm con}\kappa_{\bar\phi}}}{2\pi|P|}
=53.00259\,\mathrm{GHz},\\
\left(\frac{|\bar\phi|}{|\dR|}\right)_{\rm model}
&=\sqrt{\frac{\kappa_{\rm con}}{\kappa_{\bar\phi}}}
=1.72576843\times10^8\,\mathrm{rad/m}.
\end{aligned}
\label{eq:source_S58}
\end{equation}
The LLG time-domain fits of Sec.~\ref{sm:llg} give
\[
\begin{aligned}
f_{\rm LLG}&=49.52188\,\mathrm{GHz},\\
\frac{|A_{\bar\phi}|}{|A_{\dR}|}
&=1.57493011\times10^8\,\mathrm{rad/m}.
\end{aligned}
\]
\[
\psi_{\bar\phi}-\psi_{\dR}=90.01048^\circ.
\]
The model phase is $+90^\circ$ for the positive $P$ above.
Using the unrounded static inputs, the relative differences are
\[
\left(\frac{f_0}{f_{\rm LLG}}-1\right)=7.029\%,
\]
\[
\left[
\frac{\sqrt{\kappa_{\rm con}/\kappa_{\bar\phi}}}
{|A_{\bar\phi}|/|A_{\dR}|}-1\right]=9.577\%.
\]
The same inputs give
\[
m^*=P^2/\kappa_{\bar\phi}=1.09694654\times10^{-25}\,\mathrm{kg}.
\]
All three static inputs were extracted at $\rho=2$ only.
The equality $Z_\eta\mapsto e^{\mathrm i\beta}Z_\eta$ in Sec.~\ref{sm:llg}
shows that a uniform spin rotation changes the measured helicity by the
same angle, connecting the model coordinate to the LLG observable.

On the same sequence of fixed-$q$ states, define
\[
\mathcal J_X=\{j:|\dR_j|\le X\},\qquad
e_{u,j}\simeq c_e(X)+s_e(X)\dR_j.
\]
The linear fits give
\[
\begin{array}{c|c|r}
X\ (\mathrm{nm})&|\mathcal J_X|&s_e(X)\ (\mathrm{nm^{-1}})\\ \hline
0.40&3&-7.58\times10^{-4}\\
0.60&5&+3.39\times10^{-4}\\
0.80&7&+5.94\times10^{-4}\\
1.00&9&+5.34\times10^{-4}
\end{array}
\]
The nearest nonzero measured radius differences satisfy
\[
\min_{j:\dR_j\ne0}|\dR_j|\simeq0.2392\,\mathrm{nm}.
\]

For comparison, the quadratic static relation from main-text Eq.~(9),
using the interval-based estimate in Sec.~\ref{sm:ellipticity-coupling}, would give
\[
\Nn=-\frac{\zeta}{\kappa_{\Nn}}\dR,
\qquad -\frac{\zeta}{\kappa_{\Nn}}
\simeq-3.49\times10^{-3}\,\mathrm{nm^{-1}}.
\]
The LLG projection defined in Sec.~\ref{sm:llg} instead gives
\[
\frac{A_e^{\rm proj}}{|A_{\dR}|}
=1.236\times10^{-4}\,\mathrm{nm^{-1}},\qquad
\Delta\psi_e^{\rm proj}=-0.653^\circ.
\]
Its Fourier amplitude ratio at $f_{k_*}$ is separately
\[
\frac{A_{\Nn}^{\rm F}(f_{k_*})}{A_{\dR}^{\rm F}(f_{k_*})}
=6.98736\times10^{-5}\,\mathrm{nm^{-1}}.
\]
Within the variational model,
\[
\Nn^{\rm trial}(t)=\Nn[\mathbf m^{(q(t))}],
\]
and $P$ and $\kappa_{\rm con}$ are obtained from this same fixed-$q$
family, as specified in Sec.~\ref{sm:berry}.

\end{document}